\documentclass[aps,pra,twocolumn,superscriptaddress,nofootinbib,longbibliography]{revtex4-2}

\usepackage[T1]{fontenc}
\usepackage[utf8]{inputenc}
\usepackage{amsmath,amssymb,amsthm,mathtools,bm}
\usepackage{microtype}
\usepackage{physics}
\usepackage{hyperref}
\usepackage{graphicx}
\usepackage{xcolor}

\hypersetup{colorlinks=true,citecolor=blue,linkcolor=blue,urlcolor=blue}

\allowdisplaybreaks
\begin{document}

\title{Quantum scrambling of algebras of  observables: the \texorpdfstring{$\mathbb{Z}_2$}{Z2}-symmetric case}

\author{Paolo Zanardi}
\affiliation{Department of Physics and Astronomy, and Center for Quantum Information Science and Technology, University of Southern California, Los Angeles, California 90089, USA}
\affiliation{Department of Mathematics, University of Southern California, Los Angeles, California 90089-2532, USA}

\author{J.~Karson Lewis}
\affiliation{Department of Physics and Astronomy, University of Southern California, Los Angeles, California 90089, USA}
\begin{abstract}
We consider unitary quantum scrambling for an entire class of observable algebras $\mathcal A$ whose commutant ${\mathcal A}'=\mathbb{C}\mathbb{Z}_2$ is generated by a Hermitian involution $S$, i.e., a parity operator. We adopt two different, although related, scrambling measures: (a) the algebraic out-of-time-order correlator (A-OTOC) introduced in \cite{Andreadakis2023}, and (b) a Slater-determinant-type quantity—dubbed the Pl\"ucker fidelity—obtained by embedding operator algebras into a higher-dimensional projective operator space. Both measures admit a simple geometric interpretation in terms of distances between algebras and their dynamical images. Moreover, they can be expressed in terms of the norm of the $\mathbb{Z}_2$-symmetry-breaking component of the unitary, which is in turn controlled by the time autocorrelation function of the $\mathbb{Z}_2$ generator $S$. We derive exact expressions for the A-OTOC and the Pl\"ucker fidelity in the general case, as well as their values for a random unitary. For Hamiltonian-generated channels, we compute their long-time averages, as well as the typical values of these long-time averages for random generators $S$. Finally, we numerically explore the results of our formalism for interacting spin chains in different phases and for different physical parity operators.
\end{abstract}

\maketitle

\section{Introduction}
In this paper we continue our investigation of the notion of quantum scrambling for whole algebras of observables advocated in \cite{Zanardi2022,Andreadakis2023,Andreadakis2024,Styliaris2021}.
The idea extends the formalism of out-of-time-order correlators (OTOCs) \cite{LarkinOvchinnikov1969_QuasiclassicalMethod,ShenkerStanford2014_BlackHolesButterfly,MaldacenaShenkerStanford2016_BoundOnChaos,SwingleBentsenSchleierSmithHayden2016_MeasuringScrambling}
and starts by
associating to two distinct sets of quantum degrees of freedom --or generalized subsystems-- an operator algebra $\mathcal{A}$ and its commutant $\mathcal{A}'$.
The dynamically evolved observables associated with one set develop non-trivial commutators with those associated with the other. In this way, quantum information initially encoded in one of the two generalized subsystems can spread into the other.
This dynamical delocalization process is what we refer to as  quantum ``scrambling'' of algebras of observables.

In our approach, this type of scrambling is quantified by a coarse-grained measure of dynamically induced non-commutativity.
More precisely, an algebraic OTOC (A-OTOC) is defined as the average squared norm of commutators between
unitaries in $\mathcal{A}$ and dynamically evolved unitaries in ${\mathcal A}'$ [see Eq.~(\ref{eq:A-OTOC}) below].

Remarkably, such A-OTOCs can be cast in closed analytical form for all Hermitian-closed $\mathcal{A}$ \cite{Zanardi2022,Andreadakis2023,Andreadakis2024,Styliaris2021} and generalized to open systems
and finite temperature~\cite{Andreadakis2023,Zanardi2021,AnandZanardi2022_BROTOCs}.
The freedom in the choice of the algebra $\mathcal A$ gives the A-OTOC great flexibility, allowing it to capture aspects of different quantum-mechanical resources and, in a way, to interpolate between them.

For example, for algebras $\mathcal A$ that are \emph{factors}\footnote{Namely with trivial center, see Sect.~\ref{sec:setup}}, the corresponding A-OTOC coincides with operator entanglement \cite{Zanardi2001_EntanglementQuantumEvolutions,WangZanardi2002_QuantumEntanglementUnitaryOperators} with respect to an ordinary Hilbert space bi-partition. Operator entanglement and the related concept of entangling power \cite{ZanardiZalkaFaoro2000_EntanglingPower} can be used as powerful diagnostic tools to investigate quantum chaos, integrability, and localization in many-body systems \cite{ProsenPizorn2007_OSEE_Ising,ZhouLuitz2017_OperatorEE_ChaoticSystems,PalLakshminarayan2018_EntanglingPower_ManyBody,PizornProsen2009_OSEE_XY,MacCormackTanKudlerFlamRyu2021_NonthermalizingSystems}.

On the other hand, when the algebra is maximal abelian, it defines a complete set of commuting observables and a basis in the Hilbert space. The corresponding A-OTOC becomes identical to the coherence-generating power of the dynamics with respect to that basis \cite{ZanardiStyliarisCamposVenuti2017_CoherenceGeneratingPower}.
Such a quantity can also be used to study chaos and localization \cite{AnandStyliarisKumariZanardi2021_CoherenceSignatureChaos,StyliarisAnandCamposVenutiZanardi2019_CoherenceLocalization}, as well as  non-unitary dephasing processes \cite{StyliarisCamposVenutiZanardi2018_CoherenceGeneratingPowerDephasing}.

In both cases above, $\mathcal A$ belongs to a special class of algebras, called \emph{collinear}, for which the A-OTOC has a simple geometric interpretation in a Grassmannian manifold of operator subspaces. Specifically, the A-OTOC is the square of a natural distance function between $\mathcal{A}'$ and its dynamical image \cite{Zanardi2022,Andreadakis2023}.  Unfortunately, a metric interpretation of the A-OTOC has so far been lacking for more general algebras of observables.

In this paper we go beyond the collinear family and focus on observable algebras $\mathcal A$ whose commutant is generated by a Hermitian involution $S$, i.e., a parity operator. In this case ${\mathcal{A}}'=\mathbb{C}\mathbb{Z}_2,$ and the Hilbert space splits into invariant subspaces labelled by the $\pm 1$ eigenvalues of $S$, i.e., by parity symmetry. Under these assumptions,  the A-OTOC
and a related Slater-determinant like quantity --the Pl\"ucker fidelity-- can be expressed compactly for any  algebra and a metric interpretation recovered. 

In the unitary case, we will show that these quantities depend only on the $\mathbb{Z}_2$-symmetry-breaking component of the evolution operator, which in turn depends only on the time autocorrelation function of the $\mathbb{Z}_2$ generator. We will study their typical values for random generators $S$ and their long-time averages for unitary channels generated by a Hamiltonian. Finally, we will consider concrete spin chains to test, by numerical means,  the ability of our formalism to provide diagnostic tools for different physical phases and the transitions between them.

The paper is organized as follows. In Sec.~\ref{sec:setup} we briefly introduce the general formalism of A-OTOC and of Pl\"ucker fidelity. In Sec.~\ref{sec:general-z2}, we specialize to the general $\mathbb{Z}_2$-symmetric case, giving explicit analytical forms for our scrambling functions  and provide a metric interpretation. In Sec. \ref{sec:typ-LT}  we briefly study   their typical values and their long-time averages for Hamiltonian evolutions.  In Sec.~\ref{sec:spin-chains} we provide extensive numerical investigations of interacting spin chains of physical interest. Finally, Sec.~\ref{sec:conclu} contains the conclusions and outlook.
\section{General Setup\label{sec:setup}}
\label{sec:general-identities}
In this paper we will focus on a finite-dimensional Hilbert space $\mathcal{H},\,\mathrm{dim}\,\mathcal{H}=d.$ Its operator space $L(\mathcal{H})$ is endowed with the Hilbert-Schmidt
inner product $$\ev{X,Y}=\Tr(X^\dagger Y),\,\|X\|_2^2=\ev{X,X}.$$
The superoperator  Hilbert-Schmidt type norm is defined as $\vert|\mathcal{T}\vert|_{\mathrm{HS}}^2= \sum_{\ell,m=1}^d \|\mathcal{T}(|\ell \rangle\langle m|)\|_2^2.$

Let $\mathcal{A} \subset L(\mathcal H)$ be a $*$-algebra on the Hilbert space $\mathcal H$, and let $\mathcal{A}'$ denote its commutant. The Hilbert space decomposes as $\mathcal H \cong \bigoplus_J \mathbb{C}^{n_J} \otimes \mathbb{C}^{d_J},$ and the algebras
\begin{align}
\mathcal{A}= \bigoplus_J I_{n_J} \otimes L(\mathbb{C}^{d_J}),\quad
\mathcal{A}' = \bigoplus_J L(\mathbb{C}^{n_J}) \otimes I_{d_J}.
\end{align}
 $\mathrm{dim}\,\mathcal{A}=\sum_J d_J^2=:d(\mathcal{A}),\,\mathrm{dim}\,\mathcal{A}'=\sum_J n_J^2=:d(\mathcal{A}').$
These dimensions are related by the lower bound
\begin{equation}
d(\mathcal{A}) d(\mathcal{A}')\ge d^2
\label{eq:dAdA'}
\end{equation}
This implies that, once the dimension of the algebra is fixed, the dimension of the commutant cannot be smaller than some constant.
When (\ref{eq:dAdA'})  is saturated the algebra is said to be \emph{collinear}.  An equivalent characterization is given by 
$(n_J/d_J)^2=d(\mathcal{A}')/d(\mathcal{A}),\,(\forall J),$ i.e., the vectors $\vec{n}=(n_J)_J$ and $\vec{d}=(d_J)_J$ are collinear.

Early applications of our formalism to coherence-generating power \cite{Zanardi2018}
and scrambling in bipartite systems \cite{Styliaris2021} are based on this class of algebras. One of the goals of this paper is to go past that limitation.

A convenient operator basis for $\mathcal{A}'$ is indexed by
$\alpha,\beta\equiv (J,\ell,m)$, with $\ell,m=1,\dots,n_J$, and it is given by
\begin{align}
f_{\alpha}&={(\ket{\ell}\!\bra{m}\otimes I_{d_J})}/{\sqrt{n_J}},\quad
\tilde f_\beta= f_\beta/\|f_\beta\|_2,\\
&\ev{f_\alpha,f_{\alpha'}}=\delta_{\alpha,\alpha'}\,{d_J}/{n_J},\quad
\ev{\tilde f_\beta,\tilde f_{\beta'}}=\delta_{\beta,\beta'}.
\end{align}
Here, $\|f_\alpha\|_2^2=d_J/n_J.$
The orthogonal projectors onto $\mathcal{A}$ and $\mathcal{A}'$ are CP-maps that can be expressed as \cite{Zanardi2022}
\begin{equation}
\mathbb{P}_{\mathcal{A}}(\cdot)=\sum_\alpha f_\alpha \cdot f_\alpha^\dagger,\quad\mathbb{P}_{\mathcal{A}'}=\sum_\alpha \langle \tilde{f}_\alpha,\cdot\rangle  \tilde{f}_\alpha.
\label{eq:proj}
\end{equation}
Note, $\sum_\alpha\|f_\alpha\|_2^2=d,\,\sum_\alpha |\mathrm{Tr} f_\alpha|^2=d(\mathcal{A}).$

When $\mathcal{A}=\mathcal{A}',$ one has $d_J=n_J=1,\,(\forall J)$ and the algebra is a maximal abelian algebra, i.e., a complete set of commuting observables. Moreover, if 
$\mathcal{A}\cap\mathcal{A}'=\mathbb{C}\,\mathbb{I},$ one has just one $J,$ the Hilbert space becomes a tensor product of two factors and the algebra is called a factor.

Now we turn to metric structures. If $\mathcal{A},$ and $\mathcal{B}$ are two subalgebras in $L(\mathcal{H})$ one can define a  distance between them by using the corresponding projectors  \cite{Zanardi2022}
\begin{equation}
D_{\mathrm{HS}}(\mathcal{A}, \mathcal{B}):=\vert| \mathbb{P}_{\mathcal{A}} -\mathbb{P}_{\mathcal{B}}\vert|_{\mathrm{HS}}.\label{eq:dist-HS}
\end{equation} 
This is a first natural choice once one regards $\mathcal{A}$ and $\mathcal{B}$ as elements of the Grassmannian of subspaces of $L(\mathcal{H}).$ 
Later we will introduce other metric functions and relate them to scrambling of algebras.
\subsection{A-OTOC\label{sec:A-OTOC}}

In this section we present, following Refs.~\cite{Zanardi2022,Andreadakis2023}, the main ingredients of our paper:
the A-OTOC and the Pl\"ucker fidelity. These objects will be used to measure how the observables in the algebra $\mathcal{A}$ are, on average, scrambled by the dynamics described by a unitary channel
$\mathcal{U}(\cdot):= U\cdot U^\dagger,$ where $U$ is a unitary map on $L(\mathcal H).$ These measures carry, besides the dynamical one, an algebraic and a geometrical meaning. 

We define the A-OTOC associated with $\mathcal{A}$ and $U$ by
\begin{equation}
G_{\mathcal A}(U)
= \frac{1}{2d}\,\mathbb E_{\substack{X\in \mathcal A\\ Y\in \mathcal A'}}
\Bigl[\,\lVert [X,\mathcal U(Y)]\rVert_2^2\Bigr].
\label{eq:A-OTOC}
\end{equation}
Here $\mathbb E$ denotes a Haar average over unitaries $X\in \mathcal A$ and $Y\in \mathcal A'$. The A-OTOC measures the lack of commutativity between $\mathcal A$ and the  image of its commutant.
\begin{equation}
\mathcal U(\mathcal A'):=\{U Y U^\dagger \mid Y\in \mathcal A'\}
 \end{equation}
 The bound $G_{\mathcal A}(U)\le\mathrm{min}\{ 1-1/d(\mathcal{A}),\, 1-1/d(\mathcal{A}')\},$ holds and shows that  maximal scrambling --as quantified by the A-OTOC-- is controlled by the smallest algebra between $\mathcal A$ and its commutant.
The focus of this paper will be on smallest non-trivial commutant $d(\mathcal{A}')=2,$ thus the A-OTOC will be upper-bounded by $1/2.$
 
The A-OTOC (\ref{eq:A-OTOC})  which is defined in terms of dynamically-induced non commutativity in some cases admits a transparent metric interpretation. Indeed, we recall from \cite{Andreadakis2023} that, in the collinear case, one has  
\begin{equation}
G_{\mathcal{A}}(U)=\frac{1}{2 d(\mathcal{A}')}D^2_{\mathrm{HS}}(\mathcal{A}', \mathcal{U}(\mathcal{A}'))\label{eq:metric-coll}
\end{equation}
Namely, in this case the A-OTOC has a straightforward  geometrical meaning as the distance between the algebras $\mathcal{A}'$ and ${\mathcal{U}}(\mathcal{A}').$ 
One of the motivations of this paper is to extend the metric interpretation beyond the class of collinear algebras [See. Eqs.~(\ref{eq:z2-HS}), (\ref{eq:metric-Z2})].

Going back to the general unitary case, one finds \cite{Andreadakis2023}
\begin{align}
G_{\mathcal A}(U)=1-
\frac{1}{d}
\lVert \hat M_{\mathcal A}(U)\rVert_2^2.
\label{eq:one-minus-G}
\end{align}
Here we introduced the $d(\mathcal A')\times d(\mathcal A')$ matrix of correlators
\begin{equation}
\bigl(\hat M_{\mathcal A}(U)\bigr)_{\alpha\beta}
:=\langle \tilde f_\alpha,\mathcal U(f_\beta)\rangle.
\label{eq:Mhat}
\end{equation}
This quantity vanishes iff $\lVert \hat M_{\mathcal A}(U)\rVert_2^2=d$, which in turn holds iff
$\mathcal U(f_\alpha)=\mathbb P_{\mathcal A'}\,\mathcal U(f_\alpha),
\,(\forall\alpha),$ or in a purely super-operator form
\begin{equation}
\mathcal{U} \mathbb{P}_{\mathcal A'}= \mathbb{P}_{\mathcal A'} \mathcal{U}  \mathbb{P}_{\mathcal A'}.
\end{equation}
This is to say that $\mathcal{A}'$ --and therefore $\mathcal A$-- is invariant under the action of the channel  $\mathcal U$  \cite{Andreadakis2023}. 
This no scrambling property is clearly satisfied by any $U\in\mathcal{A}'\vee\mathcal{A}=\mathrm{span}\{ a b\mid a\in \mathcal{A},\,b\in\mathcal{A}'\}$\footnote{
In particular, for abelian commutants, $\mathcal{A}'\subset \mathcal{A}$ and this type of non-scrambling unitaries are in $\mathcal A.$}
However, besides these ``inner" non scrambling unitaries  one may also find ``outer" ones not in $\mathcal{A}'\vee\mathcal{A},$
[see e.g., Example {\bf{a)}} in Sect.~\ref{sec:general-z2}].


\subsection{Pl\"ucker fidelity}
Another geometrical object measuring the distance between the algebras $\mathcal A'$ and $\mathcal U(\mathcal A')$ can be introduced by resorting to the following construction; see Ref.~\cite{Zanardi2018}. 

Let $\mathcal B$ be a subalgebra of $\mathcal L(\mathcal H)$ with orthonormal basis $\{e_\alpha\}_{\alpha=1}^{d(\mathcal B)}$; we associate to this operator subspace the totally anti-symmetric vector in $L(\mathcal{H})^{\otimes\, d(\mathcal B)},$
\begin{align}
\Psi(\mathcal B)
:= \frac{1}{{\sqrt{d(\mathcal B)!}}}
\sum_{\sigma\in S_{d(\mathcal B)}} (-1)^{|\sigma|}
\bigotimes_{\alpha=1}^{d(\mathcal B)} e_{\sigma(\alpha)}
\label{eq:PsiB}
\end{align}
Under a basis change, this vector changes just by a scalar factor (the determinant of the base-change matrix), so its projective class $[\Psi(\mathcal A')]$ is a function of $\mathcal A'$ only. 
Now, in the projective space one can define an overlap function between two equidimensional algebras $\mathcal B$ and $\widetilde{\mathcal B},$ by
\begin{align}
F(\mathcal B,\widetilde{\mathcal B})
:= \bigl|\langle \Psi(\mathcal B),\Psi(\widetilde{\mathcal B})\rangle\bigr|
= \bigl|\det \hat O(\mathcal B,\widetilde{\mathcal B})\bigr|,
\label{eq:overlap}
\end{align}
where we have introduced a  $d(\mathcal{A}')\times d(\mathcal{A}')$
$\hat O(\mathcal B,\widetilde{\mathcal B})_{\alpha\beta}
:= \langle e_\alpha,\tilde e_\beta\rangle.
$
Once this is done, one can define distance functions between algebras, e.g.,
\begin{align}
 D_{FS}(\mathcal B,\widetilde{\mathcal B})&:=\acos F(\mathcal B,\widetilde{\mathcal B}),\nonumber\\
 D_2(\mathcal B,\widetilde{\mathcal B})&:=\sqrt{1-F^2(\mathcal B,\widetilde{\mathcal B})}.
\label{eq:dist-alg}
\end{align}
The idea here, in analogy with the A-OTOC case above, is to use these distances between $\mathcal{A}'$ and $\mathcal U(\mathcal A')$ to quantify the amount of scrambling induced by $U.$
Using these definitions we introduce the Pl\"ucker fidelity ~\cite{Zanardi2018} by
\begin{equation}
F_{\mathcal A}(U):=F(\mathcal A',\mathcal U(\mathcal A'))
= \bigl|\det \widetilde M_{\mathcal A}(U)\bigr|,
\label{eq:FA}
\end{equation}
where we have defined another $d(\mathcal A')\times d(\mathcal A')$ matrix of correlators  which is the non-trivial component of $ \mathbb{P}_{\mathcal A'} \mathcal{U}  \mathbb{P}_{\mathcal A'}$
\begin{equation}
\bigl(\widetilde M_{\mathcal A}(U)\bigr)_{\alpha\beta}
:= \langle \tilde f_\alpha,\mathcal U(\tilde f_\beta)\rangle,
\label{eq:Mtilde}
\end{equation}
Note that one has
$\widetilde M_{\mathcal A}(U)=\hat M_{\mathcal A}(U)\, Q,
$
where
\begin{equation}
Q=\operatorname{diag}\bigl(\lVert f_1\rVert_2^{-1},\ldots,\lVert f_{d(\mathcal A')}\rVert_2^{-1}\bigr),
\label{eq:Dmatrix}
\end{equation}
hence the Pl\"ucker fidelity can also be written in terms of the $\hat M_{\mathcal A}(U)$ from above as
$F_{\mathcal A}(U)= N_{\mathcal A}\,\bigl|\det \hat M_{\mathcal A}(U)\bigr|,
$
where,
$N_{\mathcal A}
= \prod_{\alpha=1}^{d(\mathcal A')} \lVert f_\alpha\rVert_2^{-1}
= \prod_J \left(\frac{n_J}{d_J}\right)^{n_J^2/2}.
$

\section{General $\mathbb{Z}_2$ theory\label{sec:general-z2}}

The focus of this paper is to study scrambling for algebras whose commutant is two-dimensional,
$\mathcal{A}' = \mathbb{C}\mathbb{Z}_2 =  \mathbb{C} \oplus \mathbb{C}.
$
In this case one has the decomposition of the Hilbert space into two orthogonal ``parity'' sectors,
$$\mathcal{H} = \bigoplus_{\alpha=\pm} \mathcal{H}_{\alpha},\quad
 \dim \mathcal{H}_{\alpha} = d_{\alpha}.$$
 Accordingly, the algebra $\mathcal{A}$ has the structure,
$$\mathcal{A} = \bigoplus_{\alpha=\pm} \operatorname{Mat}_{\mathbb{C}}(d_{\alpha}).
$$
The projectors $f_\pm$ onto $\mathcal{H}_\pm$ can be taken as a basis of $\mathcal{A}',$ however it is more convenient
to introduce a single Hermitian generator $S,$  squaring to the identity. In this way $\mathcal{H}_\pm$ are just the eigenspaces of  $S$ associated to the eigenvalues
$\pm 1.$ 
In summary,
$
\mathcal{A}' =\operatorname{span}\{\tilde f_\alpha\}_{\alpha=\pm},$ where,
\begin{equation}
\left\{
\begin{aligned}
S&=S^\dagger=S^{-1},\qquad d_\alpha = \frac{d (1+\alpha\tau)}{2}=\Tr f_\alpha,\\[3pt]
\tilde f_\alpha &= \frac{1}{\sqrt{d_\alpha}} f_\alpha := \frac{\mathbb{I}+\alpha S}{2\sqrt{d_\alpha}},\,\,(\alpha=\pm)\\[3pt]
\mathbb{I}&=f_+ +f_-,\qquad f_+f_-=0,\\[3pt]
 \tau&:=\frac{d_+-d_-}{d} = \frac{\Tr S}{d}.
\end{aligned}
\right.
\label{eq:z2-data}
\end{equation}
The collinear case corresponds to $d_+=d_-$ i.e., $\tau=0.$ 
 Moreover, the projector onto $\mathcal A$ is given by
\begin{equation}
\mathbb{P}_{\mathcal{A}}(X)=\sum_{\alpha=\pm1} f_\alpha X f_\alpha^\dagger = \frac12\qty(X+SXS).
\label{eq:PA-z2}
\end{equation}
While $d({\mathcal A}')=2$ is fixed, one has that $d({\mathcal A})=\Tr_{\mathrm{HS}} \mathbb{P}_{\mathcal{A}},$ depends on  the trace parameter $\tau$
$$d({\mathcal A})=d_+^2+d_-^2=\frac{d^2}{2} (1+\tau^2),$$ which
is minimum (maximum) for $\tau=0$ ($\tau=1$). 
Also, $|\tau|<1,$ as $\tau=\pm 1$ corresponds to $S=\pm \mathbb{I}$ in which case  there is no real $\mathbb{Z}_2$-symmetry.
The collinear case is obtained in the traceless situation $\tau=0$ i.e., when the parity sectors are isomorphic ($d_+=d_-$).  

The following function will play a key role in the rest of the paper
\begin{equation}
Y(U):=\frac{1}{d}  \langle S, {\mathcal{U}}(S)\rangle. 
\label{eq;Y}
\end{equation}
The range of $Y$ is $[2|\tau|-1,1]$ [see Eq.~(\ref{eq:Y-range})].
This  quantity  can be thought of as the infinite-temperature autocorrelation function of the $\mathbb{Z}_2$-generator $S$ with respect to the dynamics described by the channel $\mathcal U$.
Clearly, $|Y(U)|\le 1,$
and the upper-bound is saturated iff $\mathcal{U}(S)=\pm S.$  We will see that these cases correspond to the no-scrambling condition.

\subsection{Examples}
To illustrate our formalism with simple examples let us now consider a few physically motivated cases.
\begin{itemize}
\item[\bf{a)}]  For $\mathcal{H}=(\mathbb{C}^{2})^{\otimes N}$ ($N$-qubit space), any element
$P=P^\dagger$ in the Pauli group generates a $\mathbb{Z}_2$-symmetry. In this case
$\mathcal{A}'=\mathbb{C}[\mathbb{I},P]$. If $P\neq \mathbb{I}$, the symmetry is
nontrivial and the space splits as $\mathcal{H}=\bigoplus_{\alpha=\pm1}\mathcal{H}_\alpha,$ where $ \mathcal{H}_\alpha\cong(\mathbb{C}^{2})^{\otimes (N-1),}$
 correspond to the $\alpha$-eigenspace of $P$.
Set $N=1, P=\sigma^z$ and $U=\exp(-i t {\Lambda}\, \vb*{m}\cdot\vb*{\sigma}),\,\| \vb*{m}\|=1.$
If $\vb*{m}^\perp=(m_x, m_y,0)$ and $s(t):=\sin(\Lambda t),$ a simple computation yields,  
\begin{align}
Y(t)&=1-2 s^2(t) \|\vb*{m}^\perp \|^2,\nonumber \\
G_{\mathcal A}(t)&= 2 s^2(t) \|\vb*{m}_\perp \|^2 \left(1- s^2(t) \|\vb*{m}_\perp \|^2\right).\nonumber
\end{align}
For $ \|\vb*{m}_\perp \|=1,$ this simplifies to $G_{\mathcal A}(t) =\frac{1}{2}s^2(2t).$ 
This quantity vanishes for $\Lambda t= n \pi/2, (n\in\mathbb{Z})$, even/odd integers corresponds to $U\sigma^z U =\pm \sigma^z$
and to inner and outer unitaries respectively.
\item[\bf{b)}] Let $\mathcal{H}=\mathbb{C}^{d}\otimes\mathbb{C}^{d}$, and let $S$ be the swap
operator. Here $\mathcal{A}'=\mathbb{C}[\mathbb{I},S]$ is the group algebra of the
permutation group $S_2$. Now, $\mathcal{H}_\alpha\cong\mathbb{C}^{d_\alpha}$,
$d_\alpha:=d(d+\alpha)/2$, $(\alpha=\pm1)$. 
Scrambling amounts to the breaking of permutation symmetry.

\item[\bf{c)}] Let $\mathcal{H}=\mathbb{C}^{d}$ and fix $|\psi\rangle\in\mathcal{H}$. Consider, as a
$\mathbb{Z}_2$-generator, $S=2|\psi\rangle\langle\psi|-\mathbb{I}$. One has
$\mathcal{H}_{+1}\cong\mathbb{C}|\psi\rangle$,
$\mathcal{H}_{-1}=\mathbb{C}^{d-1}:=\{\,|\phi\rangle\in\mathcal{H}\mid \langle\phi|\psi\rangle=0\,\}$,
and $$1-Y=\frac{4}{d} \,(1-|\langle\psi|U|\psi\rangle|^2).$$ Here scrambling is proportional to
the infidelity of the state $|\psi\rangle$ under the unitary $U$.
\end{itemize}
Example {\bf{a)}} will be extended to spin-chains in Sect. \ref{sec:spin-chains}. Also, note that
  {\bf{b)}} is always non collinear, whereas example {\bf{c)}} it is collinear just for $d=2.$
\subsection{A-OTOC and Pl\"ucker fidelity}
Now we can explicitly compute the matrix $\hat M$ and therefore the A-OTOC,
\begin{equation}
(\hat{M}_{\mathcal A})_{\alpha\beta}=
\frac{1+(\alpha+\beta)\tau+\alpha\beta Y}
     {2\sqrt{2(1+\alpha \tau)}},
\qquad \alpha,\beta\in\{\pm1\},
\label{eq:M-z2-explicit}
\end{equation}
A straightforward computation yields the expression
\begin{equation}
d^{-1}\|\hat{M}_{\mathcal A}(U)\|_2^2 =  \frac{1+{Y}^2-2{Y}\tau^2}{2(1-\tau^2)}.
\label{eq:Mnorm-z2-closed}
\end{equation}
Furthermore, by using Eq.~\eqref{eq:M-z2-explicit}, the matrix $\widetilde M$ can be written explicitly as
\begin{equation}
(\widetilde M_{\mathcal A})_{\alpha\beta}=\frac{1+(\alpha+\beta)\tau+\alpha\beta Y}{2\sqrt{(1+\alpha \tau)(1+\beta \tau)}},
\qquad \alpha,\beta\in\{\pm1\}.\label{eq:tilde-M}
\end{equation}
One can now obtain, in this $\mathbb{Z}_2$-symmetric case,  a closed form for the A-OTOC  and the Pl\"ucker fidelity  which depends just on the parameters $\tau^2$ and $Y=Y(U),$
\begin{align}
G_{\mathcal A}(U)&=(1-Y)\qty[1-\frac{1-Y}{2(1-\tau^2)}],\nonumber \\
F_{\mathcal A}(U)&=\left|\frac{Y-\tau^2}{1-\tau^2}\right|.
\label{eq:G-F-Y}
\end{align}
The singularities for $|\tau|=1,$ are removable and the limits of  the A-OTOC and the fidelity for $|\tau|\to 1,$ are $0$ and $1$ respectively.
The functions (\ref{eq:G-F-Y}) are the main ingredients of this paper\footnote.  Note, that they are not independent as one can easily check that they satisfy $$G_{\mathcal A}(U)=\frac{1-\tau^2}{2}[1-F^2_{\mathcal A}(U)].$$
This relation vindicates the intuition that the higher the fidelity the smaller the scrambling, and vice versa.
It also shows that while  the A-OTOC  and the Pl\"ucker fidelity  are simple functions of the autocorrelation $Y,$ these functions are non-trivial e.g., non-monotonic, as they depend on the particular combination $|Y-\tau^2|.$


From Eqs. (\ref{eq:G-F-Y})  it follows that the no scrambling condition, $G_{\mathcal A}(U)=0,$ is achieved only for  

i) $Y=1,$   which holds iff ${\mathcal U}(S)=S$ i.e., $U\in\mathcal{A},$ or
\begin{equation}
\mathrm{ii)}\quad1-Y=2(1-\tau^2)\Longleftrightarrow {\mathcal U}(S)=-S
\label{eq:outer}
\end{equation}
This case corresponds to  $U$ anti-commuting with $S.$ 
Clearly, ii) can occur just for traceless $S$ and correspond to an outer automorphisms of the algebras.

In summary, the set of non-scrambling unitaries, $G_{\mathcal A}(U)=0$, is the
normalizer
\begin{align}
N(\mathcal A')
:=\{\,U\in U(\mathcal H)\mid \mathcal U(\mathcal A')=\mathcal A'\,\},
\end{align}
and has the following structure
\begin{align}
N(\mathcal A')
=
\begin{cases}
U(\mathcal A)\cong U(d_{+})\times U(d_{-}), & \text{for } |\tau|>0,\\[2mm]
U(\mathcal A)\rtimes \mathbb{Z}_2, & \text{for } \tau=0.
\end{cases}
\label{eq:normalizer}
\end{align}
where the $\mathbb{Z}_2$ is generated by the block off-diagonal operator coupling the different parity sectors
\begin{align}
R=
\begin{pmatrix}
0 & \mathbb{I}_{d/2}\\
\mathbb{I}_{d/2} & 0
\end{pmatrix},
\end{align}
such that
$R(U_1\oplus U_2)R^{-1}=U_2\oplus U_1,
\;
RSR^\dagger=-S.
$

We now move to consider the opposite situation of maximal scrambling. From Eq.~(\ref{eq:G-F-Y}) it follows that  the following upper-bound holds for all $U$'s
\begin{equation}
G_{\mathcal A}(U)\le \frac{1-\tau^2}{2}=:G_{\mathrm{max}}(\tau).\label{eq:G-upper-bound}
\end{equation}
Such a maximum is achieved for unitaries such that $Y(U)=\tau^2,$ or equivalently, $F_{\mathcal{A}}(U)=0.$ 
By a Haar average over the $U$s one finds 
$\langle U SU^\dagger\rangle_U =\mathbb{I} \Tr(S)/d,$ hence
\begin{align}
\langle Y(U)\rangle_U&=\tau^2.\label{eq:Y-ave-U}
\end{align}
Since $Y(U)$ is a continuous function and $U(\mathcal{H})$ is connected it follows that maximally scrambling unitaries satisfying $Y(U)=\tau^2,$ always exist.  
This existential proof can be made constructive. Let us illustrate it for  traceless $S$ i.e.,  
$\tau=0$. Indeed, in this case
$\mathcal H=\mathbb C^{d/2}\oplus\mathbb C^{d/2}
\simeq \mathbb C^2\otimes\mathbb C^{d/2}$ and
$S=\sigma^z\otimes\mathbb I_{d/2}$. 
Let
$$U_t=\cos t\,\mathbb I+i\sin t\,\sigma^x\otimes C
=\exp(i t\,\sigma^x\otimes C),$$ where
 $C$ is a Hermitian
involution on $\mathbb C^{d/2}.$ 
Using $\{\sigma^x,\sigma^z\}=0$ and
$\sigma^x\sigma^z=-i\sigma^y$, one obtains
$U_tS U^\dagger_t
=\cos(2t)\,\sigma^z\otimes\mathbb I_{d/2}
+\sin(2t)\,\sigma^y\otimes C .
$
At $t=\pi/4$, $\widetilde S:=U({\frac{\pi}{4}})S U^\dagger({\frac{\pi}{4}})
=\sigma^y\otimes C$. Hence
\begin{equation}
Y(U_{\frac{\pi}{4}})=d^{-1}\operatorname{Tr}(S\widetilde S)
=d^{-1}\operatorname{Tr}\!\left[(\sigma^z\sigma^y)\otimes C\right]=0,
\end{equation}
because $\operatorname{Tr}(\sigma^z\sigma^y)=0$. This gives
$G_A(U_{\pi/4})=1/2=G_{\max}$, with the explicit maximizing unitary
\begin{equation}
U_{\frac{\pi}{4}}=\frac{1}{\sqrt2}\left(\mathbb I+i\sigma^x\otimes C\right),
\end{equation}
for any Hermitian involution $C$ on $\mathbb C^{d/2}$.

\subsection{Symmetry breaking parameter}
From its very definition Eq.~(\ref{eq:A-OTOC}) it is clear that  our measure of scrambling is tightly related to the notion of symmetry-breaking induced by the dynamics.  If $[U,S]=0,$ clearly the A-OTOC vanishes.  No scrambling, however, does \emph{not} require exact symmetry preservation.  For example, if $\{U, S\}=0,$ symmetry is maximally broken by the dynamics, but $\mathcal{A}'=\mathrm{span}\{\mathbb{I}, S\}$ is still mapped onto itself and no algebra scrambling occurs.

In the light of these considerations and to cast Eqs.~(\ref{eq:G-F-Y}) into an even more transparent form, it is useful to introduce a measure of how much  $U$ breaks the $\mathbb{Z}_2$-symmetry. Using Eq.~(\ref{eq:proj}) one finds,
\begin{equation}
\mathfrak{b}^2(U):=\frac{1}{d} \norm{{U}-\mathbb{P}_{\mathcal{A}}({U})}^2_2 = \frac12\qty(1-Y(U)).
\label{eq:D2-z2-Y}
\end{equation}
When $\mathfrak{b}(U)=\|\mathbb{P}_{{\mathcal A}^\perp}(\frac{U}{\sqrt{d}})\|_2$ vanishes, as we have seen, there is no scrambling as $\mathcal A$ and ${\mathcal A}'$ are mapped onto themselves by the channel $\mathcal U$.
For the non-collinear case $\tau\neq 0,$  in the light of (\ref{eq:normalizer}), $\mathfrak{b}(U)= 0,$ is also necessary condition for the absence of scrambling.

Interestingly,  the Haar average of the symmetry-breaking parameter leads to a simple ``probabilistic" interpretation as well as well as to an insight into the meaning of collinearity.
From Eq.~(\ref{eq:Y-ave-U}) it immediately follows that
\begin{equation}
\langle \mathfrak{b}^2(U)\rangle_U=G_{\mathrm{max}}(\tau)=1-\frac{d({\mathcal A})}{d^2}
\label{eq:D-ave}.
\end{equation}
The last equality stems from $d({\mathcal A})=d_+^2+d_-^2$ and $d\,\tau=d_+-d_-,$ from which $d({\cal A})=d^2(1+\tau^2)/2.$
Eq.~(\ref{eq:D-ave}) ranges from zero for $|\tau|\to 1,$ (no scrambling) to $1/2$ for traceless generators $S$, i.e., $d_+=d_-.$

Loosely speaking, the RHS of   (\ref{eq:D-ave})  is the ``probability'' that a random operator in $L({\cal H})$ falls into the orthogonal complement of $\mathcal A.$ 
Such random operators are in fact obtained by the action on a generic $U$ on elements of $\mathcal A;$ therefore it is intuitive that Eq.~(\ref{eq:D-ave}) plays a role
in algebra scrambling, which is based on the idea of operator algebras being dynamically evolved into orthogonal degrees of freedom.
In particular, since the collinear case ($\tau=0$), corresponds to the smallest $d(\mathcal{A})=d^2/2,$ one expects these algebras to be the most scrambling. 

Another, more algebraic, way to interpret Eq.~(\ref{eq:D-ave}) is that $1-\tau^2=d^{-1}\|S-\tau\mathbb{I}\|_2^2,$ which implies that the smaller $\tau,$ the larger the components orthogonal to the identity
of the operator in $\mathcal{A}'$ in (\ref{eq:A-OTOC}). Since those are the only components relevant to the commutators, one expects a larger A-OTOC. Again, this points to the collinear case as the most scrambling.

One can also re-express our scrambling measures as polynomial functions  of the symmetry-breaking  function $\mathfrak{b}^2$ and its Haar average,
\begin{align}
G_{\mathcal A}(U)&= {2\mathfrak{b}^2(U)}\qty(1-\frac{\mathfrak{b}^2(U)}{2 \langle \mathfrak{b}^2(U)\rangle_U}) ,\nonumber \\
F_{\mathcal A}(U)&= \left|1-\frac{\mathfrak{b}^2(U)}{\langle \mathfrak{b}^2(U)\rangle_U}\right|.
\label{eq:G-F-D}
\end{align}
Also, $G_{\mathcal A}(U)\le \langle \mathfrak{b}^2(U)\rangle_U.$
The no-scrambling conditions now read as : i) $\mathfrak{b}^2(U)=0$ (inner unitaries) and ii) $\mathfrak{b}^2(U)=2 \langle \mathfrak{b}^2(U)\rangle_U$(outer unitaries, only  for $\tau=0$).

Yet another insightful way of writing the A-OTOC can be obtained by defining a new quantity $T(U),$ the average transition probability of pure states from ${\mathcal H}_+$ to  ${\mathcal H}_-,$
and then and relating it to the symmetry breaking measure
\begin{align}
 T(U)
 &:=\mathbb{E}_{\ket{\psi}\in{\mathcal H}_+}
 \left[
 \Tr\left(f_-\,{\mathcal U}\left(\ket{\psi}\!\bra{\psi}\right)\right)
 \right]
 \notag\\
 &=\frac{1}{d_+}\Tr\left[f_-\,{\mathcal U}(f_+)\right],\nonumber\\
 \mathfrak{b}^2(U)&=(1+\tau)\,T(U).
 \label{eq:T-def}
\end{align}
Using this remark and Eq.~\eqref{eq:G-F-D},  one gets
\begin{equation}
 G_{\mathcal A}(U)
 =2(1+\tau)T(U)
 \left[
 1-\frac{T(U)}{1-\tau}
 \right].
 \label{eq:GA-transition}
\end{equation}
The no-scrambling conditions a) and b) corresponds now to $T(U)=0,$ and $T(U)=1,\,(\tau=0),$ respectively.

Eqs.~(\ref{eq:G-F-Y}), (\ref{eq:G-F-D}), and (\ref{eq:GA-transition}) are mathematically equivalent forms showing that the A-OTOC can be encoded in an autocorrelation function, a symmetry-breaking term, or dynamical transition probabilities. These views provide different physical perspectives from which one can usefully think about the scrambling of $\mathbb{Z}_2$-symmetric algebras.
\subsection{Metric structures}
One of the goals of this paper is to show  that the A-OTOC, which \emph{per se} is a measure of the algebraic property of (dynamically-induced) non-commutativity,  is directly related to metric structures in  the manifold  $\mathcal{M}_{\mathbb{Z}_2}$  of algebras with $\mathbb{Z}_2$-commutant.   Intuitively, one would like to establish a direct connection between our measures of quantum scrambling and the geometric distance by which an algebra is displaced from itself by quantum dynamics. The greater the distance the greater the scrambling.

The manifold $\mathcal{M}_{\mathbb{Z}_2}$  consists of  all
$\mathcal{A}_{S}=\operatorname{span}\{\mathbb{I},S\}\subset {L}(\mathcal{H})$,
where $S=S^\dagger$, $S^2=\mathbb{I}$ (Hermitian involution). Since $S$ and $-S$
define the same algebra, one has that $\mathcal{M}_{\mathbb{Z}_2}$ is a quotient
space,
\begin{align}
\mathcal{M}_{\mathbb{Z}_2}
\cong
\left\{
S\in {L}(\mathcal{H})\mid S=S^\dagger,\; S^2=\mathbb{I}
\right\}\big/\sim .
\end{align}
where
$S'\sim S
\;\Longleftrightarrow\;
\exists\,\alpha\in\{-1,+1\}\ \text{such that}\ S'=\alpha S .
$
Equivalently, one could consider the set of all subspaces $V$ of
$\mathcal{H}$ divided by the $\mathbb{Z}_2$-action $V\mapsto V^\perp$, or, in terms of
projectors, $P_V\mapsto \mathbb{I}-P_V$. Since $P_V=(\mathbb{I}+S)/2,$ this amounts to $S\mapsto -S$

Let $S_i,\,(i=1,2)$ be two hermitian involutions on $\mathcal{H}$ and $\mathcal{A}_{S_i}=\mathrm{span}\{\mathbb{I}, S_i\}$ the corresponding $\mathbb{Z}_2$-algebras.  An orthonormal basis for $\mathcal{A}_{S_i},$ is given by
\begin{equation}
 e_{i,0}:=\frac{\mathbb I}{\sqrt d},
 \qquad
 e_{i,1}:=\frac{S_i-\tau_i\mathbb I}
 {\sqrt{d(1-\tau_i^2)}},\nonumber
\end{equation}
where  $\tau_i:=d^{-1}\operatorname{Tr}S_i.$
In terms of this basis, the overlap matrix becomes $\hat{O}(\mathcal{A}_{S_1},\mathcal{A}_{S_2})=\mathrm{diag}( 1, \widetilde{Y}),$ where
\begin{equation}
\widetilde{Y}
:=
\frac{Y-\tau_1\tau_2}
{\sqrt{(1-\tau_1^2)(1-\tau_2^2)}},\quad Y:=\frac{1}{d}\langle S_1, S_2\rangle.
\end{equation} 
Also, the orthogonal projector onto $\mathcal B_i$ can be written as
$\mathbb P_{\mathcal B_i}=
\sum_{a=0,1}|e_{i,a}\rangle\langle e_{i,a}|.
$
Whence,
\begin{align}
D_{\mathrm{HS}}^2(\mathcal{A}_{S_1},\mathcal{A}_{S_2})&=
4-2\operatorname{Tr}_{\mathrm{HS}}
(\mathbb P_{\mathcal{A}_{S_1}}\mathbb P_{\mathcal{A}_{S_2}})\nonumber\\
&=
4-2\|\hat O(\mathcal{A}_{S_1},\mathcal{A}_{S_2})\|_2^2=2(1-\widetilde Y^2).\nonumber
\end{align}
Or,  more explicitly,
\begin{equation}
 D_{\mathrm{HS}}^2(\mathcal{A}_{S_1},\mathcal{A}_{S_2})
=2\left[
1-
\frac{(Y-\tau_1\tau_2)^2}
{(1-\tau_1^2)(1-\tau_2^2)}\right]
.\label{eq:HS-dist-Z_2}
\end{equation}
Also, $F(\mathcal{A}_{S_1},\mathcal{A}_{S_2})=|\det \hat O(\mathcal{A}_{S_1},\mathcal{A}_{S_2})|=|\widetilde Y|,$ and
$$D_2(\mathcal{A}_{S_1},\mathcal{A}_{S_2})=\sqrt{1-|\widetilde Y|^2}=\frac{1}{\sqrt{2}}D_{\mathrm{HS}}(\mathcal{A}_{S_1},\mathcal{A}_{S_2}).$$
The invariance of these relations under $Y\mapsto -Y, \tau_1\tau_2\mapsto -\tau_1\tau_2,$ shows that they are genuine functions on
$\mathcal{M}_{\mathbb{Z}_2}.$

In order to connect this to scrambling we now specialize to $\mathcal{A}_{S_1}=\mathcal{A}',\,  \mathcal{A}_{S_2}=\mathcal{U}(\mathcal{A}').$
Using Eqs.~(\ref{eq:dist-alg}),(\ref{eq:tilde-M}), and (\ref{eq:G-F-Y}) one gets that $D_{\mathrm{HS}}(\mathcal{A}', \mathcal{U}(\mathcal{A}'))=\sqrt{2} \,D_{2}(\mathcal{A}', \mathcal{U}(\mathcal{A}')),$ and 
\begin{align}
G_{\mathcal A}(U)=\frac{1-\tau^2}{4} \,D^2_{\mathrm{HS}}(\mathcal{A}', \mathcal{U}(\mathcal{A}')).
\label{eq:z2-HS}
\end{align}
Notice, the prefactor is $\frac{1}{2} \langle\mathfrak{b}^2(U)\rangle_U.$

Alternatively, since $D_2^2=\sin^2 D_{FS},$ one can write
\begin{align}
{G_{\mathcal A}(U)}
=\frac{1-\tau^2}{2}\, \sin^2 D_{FS}({\mathcal A}',\,{\mathcal U}({\mathcal A}')). \label{eq:metric-Z2}
\end{align}
Eq.~(\ref{eq:metric-coll}) is recovered from (\ref{eq:z2-HS}) for the collinear case $\tau=0.$ For non traceless $S$, the RHS of Eq.~(\ref{eq:metric-coll}) is just an upper bound to the A-OTOC.

Interestingly,  there is  another way to equip the manifold $\mathcal{M}_{\mathbb{Z}_2}$  with a natural metric structure and then relate it to
scrambling.  One can indeed  use a  quotient metric
\begin{align}
D^2_{\mathbb{Z}_2}\!\left(\mathcal{A}_{S_1},\mathcal{A}_{S_2}\right)
&:=\frac{1}{2d}\min_{\alpha,\beta=\pm1}\|\alpha S_1-\beta S_2\|_2^2 \nonumber\\
&=1-\left|\langle S_1,S_2\rangle\right| .
\end{align}
In particular,
$D^2_{\mathbb{Z}_2}\!\left(\mathcal{A}_{S},\mathcal{U}(\mathcal{A}_{S})\right)
=1-|Y({U})| .$ Since, from (\ref{eq:G-F-Y}), on has that $G_{\mathcal A}\le (1-Y^2)/2\le 1-|Y|,$ we have the upper bound
\begin{equation}
G_{\mathcal A}(U)\le D^2_{\mathbb{Z}_2}\!\left(\mathcal{A}',\mathcal{U}(\mathcal{A}')\right)
\end{equation}
This equation, along with Eqs.~(\ref{eq:z2-HS}) and (\ref{eq:metric-Z2}), show how  scrambling of algebras is quantified precisely by how much the whole set of symmetries ${\mathcal A}'$ is sent far from itself by $\mathcal{U}$ with respect to different metric structures on   $\mathcal{M}_{\mathbb{Z}_2}.$
This observation provides these, apparently abstract, metric structures with an operational meaning, and makes them potentially measurable. 
\section{Typical values\label{sec:typ-LT}}
We have  established the key properties of our scrambling functions  (\ref{eq:G-F-Y}) and their different interpretations.  it is now useful to inquire about the typical values of the A-OTOC and Pl\"ucker fidelity for random unitaries, hamiltonian flows and random $\mathbb{Z}_2$-generators. These values will provide a reference baseline for applications to concrete quantum many-body systems e.g.,  quantum spin chains.

Let us estimate  the typical value of the autocorrelation $Y(U)$ for a random generator $S.$ To this end, let us consider the ${\mathcal A}_W\cong\mathbb{C}\mathbb{Z}_2$ generated
by $S_W= W S_0 W^\dagger$ where $S_0=S_0^\dagger,\, S_0^2=\mathbb{I},\, \Tr S_0:=d\tau,$ is fixed and $W$ is Haar distributed. If $Y_W(U):=\langle S_W, {\mathcal{U}}(S_W)\rangle,$ then
\begin{align}
\langle Y_W(U) \rangle_W&=\tau^2+(1-\tau^2)\,K(U)\nonumber\\
K(U)&:=\frac{|{\Tr(U)}|^2-1 }{d^2-1}
\label{eq:Y-W-ave}
\end{align}
The quantity $K(U)$ is a kind of normalized spectral factor of $U.$  It attains its maximum $K(U)=1$ only for $U=\mathbb{I},$ and vanishes for Haar random $U$\footnote{As $\langle |{\Tr(U)}|^2\rangle_U=1$.
Also, $\langle K^2(U)\rangle_U=(d^2-1)^{-2}.$}
By concavity $$\langle G_{\mathcal{A}}(Y_W)\rangle_W\le  G_{\mathcal{A}}(\langle Y_W\rangle_W)=G_{\mathrm{max}}(\tau)[1-K^2(U)].$$

One can also think of the A-OTOC and Pl\"ucker fidelity as random variables over the unitary group of the $U$'s endowed with the Haar measure.

For large dimension $d$, one expects the typical values to be concentrated around their expectations.  Indeed, the A-OTOC is a Lipschitz function of $U,$ see Eq.~(\ref{eq:G-Lip}), and 
therefore exponential concentration results can be proved using Levy Lemma \cite{Ledoux2001Concentration}.  Here below,  however,  we will just  
use a simple Markov inequality.

The Haar average of the A-OTOC over the unitary $U$ is given by
\begin{align}
\langle G_{\mathcal A}(U)\rangle_U&=G_{\mathrm{max}}(\tau) \,\frac{d^2-2}{d^2-1}.\label{eq:conc-G}
\end{align}
For large dimension $d$ this result tends  to the upper-bound (\ref{eq:G-upper-bound})  showing that for random unitaries, scrambling is nearly maximal. More precisely, from (\ref{eq:conc-G}), using Markov inequality one gets
\begin{equation}
\mathrm{Pr}_{U\sim\mathrm{Haar}}  \{ U\mid  G_{\mathcal A}(U) \le G_{\mathrm{max}}(\tau)(1-\epsilon)\} \le \frac{2}{d^2 \epsilon}\nonumber.
\end{equation}
Setting e.g., $\epsilon=1/d,$ we see  that the RHS is $O(1/d)$ proving concentration.
Since $G_{\mathrm{max}}(\tau)$ is the largest  i.e., $1/2,$  for  $\tau=0,$  collinear algebras, as we anticipated, typically  suffer  the greatest scrambling.
%

The absolute vale in the definition of the  Pl\"ucker fidelity  (\ref{eq:G-F-Y}) makes it difficult to find a simple expression for its Haar average.
However, one  can prove  the following $\tau$-independent bound
\begin{align}
\langle F_{\mathcal A}(U)\rangle_U\le\frac{1}{\sqrt{d^2-1}}=O(1/d).
\label{eq:conc-F}
\end{align}
\subsection{Long-time averages}
In order to apply the abstract formalism presented in this paper to many-body quantum systems, we need to focus on Hamiltonian flows, namely on the case in which the dynamics is described by a unitary family $\{U(t):=e^{-iHt}\}_{t\in\mathbb{R}}$, where $H$ is the system Hamiltonian. 

A first natural approach, both mathematically and physically, is to consider the long-time averages of the A-OTOC and related functions.
We define the long-time average of a function $f$ in the standard way by
\begin{equation}
\overline{f}^{\,\infty}=\lim_{T\to\infty}\frac{1}{T} \int_0^t d\tau\, f(\tau).
\end{equation}
The basic function is
$Y(t)=\frac{1}{d} \operatorname{Tr}\!\bigl[S U_t S U_t^{\dagger}\bigr],
$
where
$U_t=e^{-iHt}=\sum_m \Pi_m e^{-iE_m t},
$
with $m\neq n \Rightarrow E_n\neq E_m$, $\Pi_n\Pi_m=\delta_{nm}\Pi_n$, and $\sum_n \Pi_n=\mathbb{I}$.
The long-time average is given by
\begin{align}
\overline{Y}^{\infty}=\frac{1}{d} \lVert D_H(S)\rVert_2^2, \label{eq:Y-ave}
\end{align}
in which $D_H(\cdot)=\sum_n \Pi_n\, \cdot\, \Pi_n,$
is the Hamiltonian dephasing superoperator.

To gain some intuition about  $\overline{Y}^{\infty}$ let us consider again the random generator case,
By setting $U=\sum_n \Pi_n e^{-i E_n t}$ in Eq.~(\ref{eq:Y-W-ave}), and if $\vec{d}=(d_n:=\Tr \Pi_n)_n,$ denotes the vector of degeneracies, the long-time average gives,
\begin{equation}
\overline{\langle Y_W\rangle}^\infty_W=1-\frac{1-\tau^2}{1-d^{-2}} \,S_l(\vec{d}), \label{eq:ave-W}
\end{equation}
where $S_l(\vec d):=1-\sum_n (d_n/d)^2=
1-d^{-2} \overline{|\operatorname{Tr} U_t |^2 }^\infty.$
This is an entropic quantity related to the degeneracy structure of $H$, and it is largest, i.e., $S_2^{\mathrm{max}}=1-1/d,$ for non-degenerate $H.$
In this case $\overline{\langle Y_W\rangle}^\infty_W=\tau^2+O(1/d),$ which is to say that the autocorrelation function is close to the value corresponding to maximal scrambling up to corrections vanishing for large $d.$

Let us now move to consider the long-time average of the A-OTOC itself. From Eq.~(\ref{eq:G-F-Y}) one obtains \footnote{Since the range autocorrelation function is given by $Y\in[2|\tau|-1,1],$ one has that  $\sigma_\infty^2(Y)\le 4(1-|\tau|)^2,$ therefore the fluctuation term in (\ref{eq:bar{G}}) which seemingly diverges for $|\tau|\to1$--  is bounded by $2(1-|\tau|).$}
\begin{equation}
\overline{G_{\mathcal {A}}(Y)}^\infty=  G_{\mathcal {A}}(\overline{Y}^{\infty}) -\frac{\sigma_\infty^2(Y)}{2(1-\tau^2)}, 
\label{eq:bar{G}}
\end{equation}
where $\sigma_\infty^2(Y)=\overline{Y^2}^{\infty}-\bigl(\overline{Y}^{\,\infty}\bigr)^2,$ is the temporal variance of $Y.$
A standard calculation based on the non-resonance condition on the spectral gaps, gives 
\begin{align}
\sigma_\infty^2(Y)=\frac{1}{d^2}\sum_{n\neq m}\bigl|\operatorname{Tr}(S\Pi_n S\Pi_m)\bigr|^2. \label{eq:t-var}
\end{align}
More generally,  if the non-resonance condition fails e.g., for free-systems, one has that $\sigma^2_\infty(Y)=\sum_{|\omega|>0} Y_\omega^2,$ where $ Y_\omega:=d^{-1} \sum_{E_n-E_m=\omega} \Tr[ S\Pi_nS\Pi_m].$

 Clearly, when the Hamiltonian commutes with the generator $S$  one has $D_H(S)=S,$ and $\bar{Y}^\infty=1,\,\sigma_\infty=0.$ On the other hand if $S$ has no non-vanishing matrix elements inside the eigenspaces of $H$\footnote{If $\Pi_n S\Pi_n=0,\,(\forall n)$ then $\tau=d^{-1}\sum_n\Tr(\Pi_n S\Pi_n)=0.$}, one has
 that $\bar{Y}^\infty=0,$ and then the long time-average of the A-OTOC is $1/2(1-\sigma_\infty^2(Y)).$ 
 Generically $\sigma_\infty^2(Y)=O(1/d^2),$ (as $|\langle m|S|n\rangle|^2\approx d^{-1}$). Hence for generators $S$ connecting only  different eigenspaces of $H$, from (\ref{eq:bar{G}}) one gets that  for large $d$  scrambling  is maximal  up to small corrections
$$
\overline{G_{\mathcal {A}}}^\infty=\frac{1}{2}-O(1/d^2).
$$ 
This is the physically important situation when  $S$ breaks some symmetries of a non-degenerate  Hamiltonian, forcing the diagonal elements $\langle n|S|n\rangle=0.$ 

\section{$\mathbb{Z}_2$-scrambling in spin chains\label{sec:spin-chains}}

In this section, we numerically study the A-OTOC in many-body systems exhibiting a variety of phenomena.  In each case, we calculate the long-time average of the A-OTOC $\overline{G(t)}^\infty$ for a parity operator $S$ which has some physical or structural relevance to the system/phenomena being studied.  All of the parity operators used in this section are traceless, so the long time average of the A-OTOC takes the form $\overline{G(t)}^\infty = \frac{1}{2}(1-\overline{Y(t)^2}^\infty)$.  For a traceless parity operator, the Haar average of the A-OTOC is $\overline{G(U)}^U\approx \frac{1}{2}$ independent of the specific choice of parity operator.  Thus, in chaotic many-body systems, the long-time average A-OTOC for a generic parity operator will be very close to its ergodic value of $\approx\frac{1}{2}$, so departures of $\overline{G(t)}^\infty$ from this value as we vary system parameters signal that the system is undergoing some structural change.  The following three examples illustrate this general behavior for different spin-chain systems.

\subsection{Many-body localization}
\begin{figure*}[t]
  \centering
  \includegraphics[scale = 0.44]{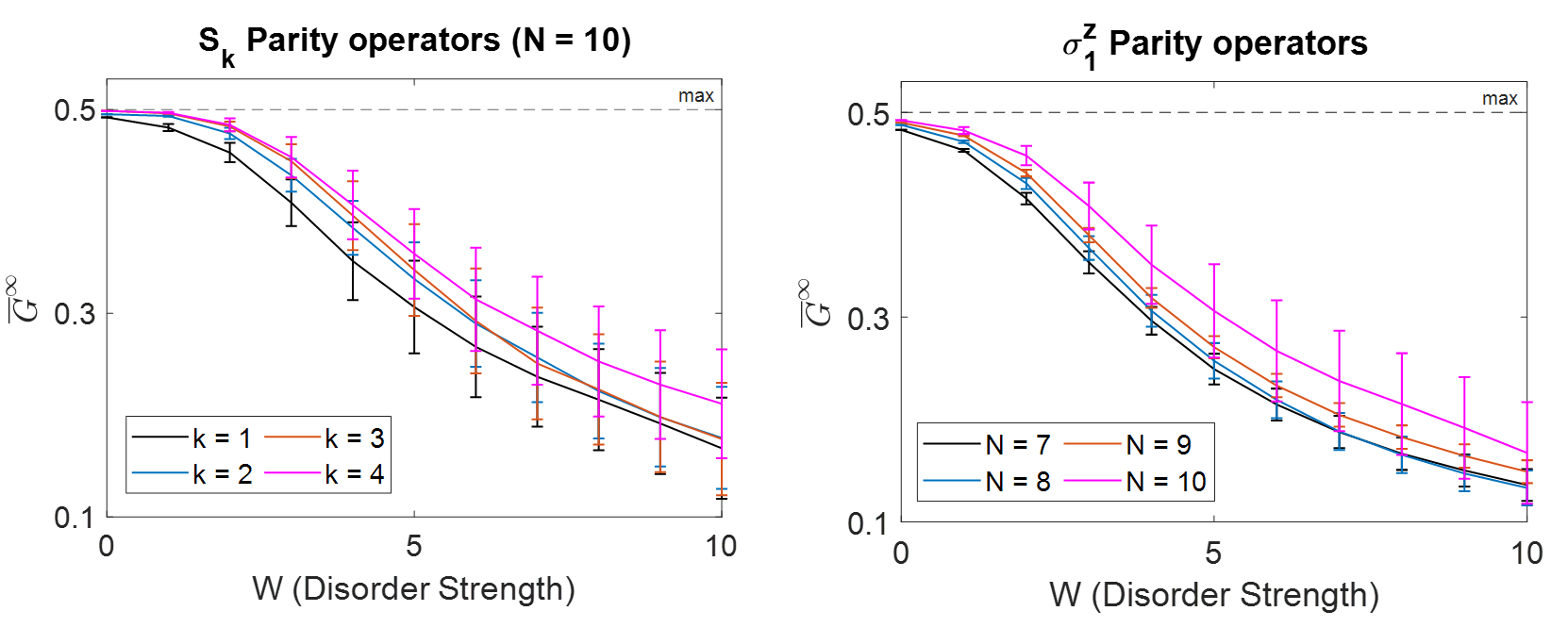}
  \caption{The left panel shows the disorder averaged $\overline{G}^\infty$ for parity operators with varying support $S_k=\prod_{i=1}^k\sigma^z_k$ for the random-field XXZ chain (\ref{eq:xxz-chain}) with $N =10$; the results are averaged over 20 realizations of the i.i.d. fields and the error bars are plotted to 2 standard errors.  For every support size tested, the A-OTOC $\overline{G}^\infty$ decreases significantly from its ergodic value $\approx\frac{1}{2}$ as the disorder increases.  The right panels shows the disorder averaged $\overline{G}^\infty$ for the parity operator $\sigma^z_1$ for different system sizes.}
  \label{fig:xxz}
\end{figure*}

Consider the random-field XXZ chain
\begin{equation}
  H=\sum_{i=1}^{N-1}\bigl(\sigma^x_i\sigma^x_{i+1}+\sigma^y_i\sigma^y_{i+1}
   +\Delta\,\sigma^z_i\sigma^z_{i+1}\bigr)+\sum_{i=1}^N\varepsilon_i\sigma^z_i,
  \label{eq:xxz-chain}
\end{equation}
with i.i.d.\ fields $\varepsilon_i$ chosen from the uniform distribution with center 0 and width $W$.  This model exhibits Anderson localization at any level of disorder for $\Delta = 0$ and many-body localization (MBL) at strong disorder for $\Delta = 1$ \cite{PhysRevLett.111.127201,pawlik2026facetsbodylocalization}.  Focusing on the MBL case, we calculate $\overline{G}^\infty$ for a family of parity operators constructed from local Pauli $z$ operators (i.e. $S_k:=\prod_{i=1}^k\sigma^z_i$); the results are shown in the left panel of Fig.~\ref{fig:xxz} for a system of 10 qubits.  We see that for all of these parity operators, $\overline{G}^\infty$ averaged over the disorder shows a significant decrease as the disorder strength is increased and the system transitions into the MBL phase.   

\begin{figure}[t]
  \centering
  \includegraphics[width=\columnwidth]{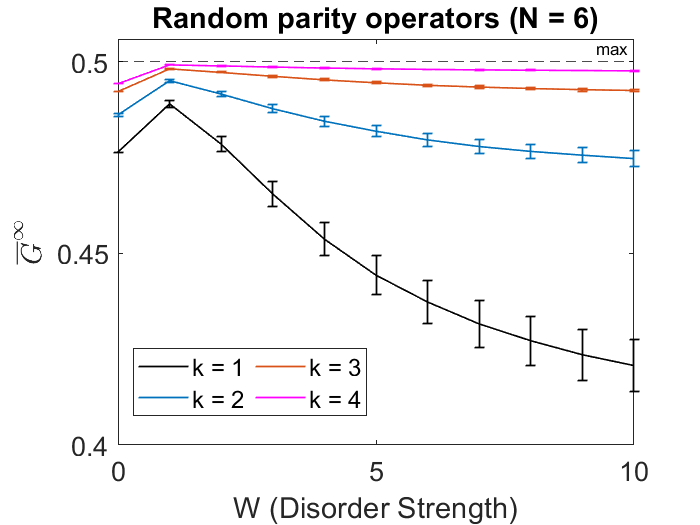}
  \caption{Disorder averaged $\overline{G}^\infty$ averaged over random parity operators of varying support.  The parity operators are given by $S = Vs_AV^\dagger\otimes I_{B}$ where $s_A:=\textnormal{diag}(+1^{\times 2^{k-1}},-1^{\times 2^{k-1}})$, $A$ is the subset of the spin-chain consisting of the left-most $k$ spins, and $B$ is the complement of $A$.  Here the average is calculated by computing $\overline{G}^\infty$ for 1000 pairs of disordered fields and Haar random unitaries $V$.}
  \label{fig:MBL2}
\end{figure}

We also see that the largest decrease in $\overline{G}^\infty$ is observed when the parity operator has the smallest support, and the parity operators typically become less sensitive to MBL as their support increases.  To see the effect of support on the ability of $\overline{G}^\infty$ to detect MBL, we choose parity operators randomly via the Haar ensemble while varying their support ($S = Vs_AV^\dagger\otimes I_{B}$ where $s_A:=\textnormal{diag}(+1^{\times 2^{k-1}},-1^{\times 2^{k-1}})$, $V$ is a Haar random unitary, $A$ is the subset of the spin-chain consisting of the left-most $k$ spins, and $B$ is the complement of $A$).  The results are shown in Fig.~\ref{fig:MBL2}.  Here we see that even these randomly chosen parity operators are able to resolve the localization of the system as the disorder strength increases, and that their ability to do so decreases with their support.  This suggests that the behavior of $\overline{G}^\infty$ through the MBL transition is somewhat robust to the choice of \emph{local} parity operators, but more sensitive to the choice of those with non-local support.

Apart from this general behavior, there are a few differences between the results for the $S_k$ parity operators and the random ones. The random parity operators experience an intial jump in the disorder-Haar averaged $\overline{G}^\infty$ shown in Fig.~\ref{fig:MBL2}; this can be seen from $\sigma^2_\infty(Y)=\sum_{|\omega|>0}Y^2_\omega$
since at zero disorder, the spectrum has degenerate gaps, but even at small disorder, the gaps become non-degenerate.  Thus, at zero disorder, $Y_\omega$ contains multiple terms, but once the disorder is turned one, all but one of these terms vanish.  $Y_\omega$ is non-negative, so the result is that $\sigma^2_\infty$ (and thus $\overline{Y^2}^\infty$) decreases abruptly when the disorder is turned on, resulting in the jump in $\overline{G}^\infty$.  In general, the jump is dependent on the choice of parity operator (e.g. it is not see with $S_k$ parity operators).  Further, the error bars shrink with increasing support for random parity operators but remain approximately the same size for the $S_k$ parity operators; this is because $\overline{G}^\infty$ is a Lipschitz function of the Haar random unitaries used to generate the random parity operators, so the shrinking of the error bars is a result of measure concentration.

The right panel in Fig.~\ref{fig:xxz} shows the disorder averaged $\overline{G}^\infty$ for the parity operator $\sigma^z_1$ for several different system sizes.  Although the behavior is consistent over all system sizes tested, we fail to see a consistent cross-over at some critical level of disorder.



\subsection{Ordered - disordered transition}

Consider the transverse-field Ising model (TFIM)
\begin{equation}
    H=-J\sum_{i=1}^{N-1}\sigma^z_i\sigma^z_{i+1}-h\sum_{i=1}^N\sigma^x_i.\label{eq:TFIM}
\end{equation}
At $h=0$, this model exhibits long-range order where all of the spins are aligned in the ground state (i.e. $\langle0|\sigma^z_i|0\rangle=1$ for all $i$ where $|0\rangle$ is the ground state).  As $h$ is increased from 0, the model undergoes a phase transition where the ground state becomes disordered.  In the thermodynamic limit, the single-site magnetization in the $z$ direction decays from 1 to 0 as $h$ increases from zero to a critical value of $h = J$; the decay is given by $\langle0|\sigma^z_i|0\rangle=(1-(\frac{h}{J})^2)^{1/8}$ and for $h\geq J$, $\langle0|\sigma^z_i|0\rangle$ remains zero \cite{Pfeuty:1970qrn}.  Thus, the single-site magnetization shows how the transverse field induces disorder in the model.

\begin{figure}[t]
  \centering
  \includegraphics[width=\columnwidth]{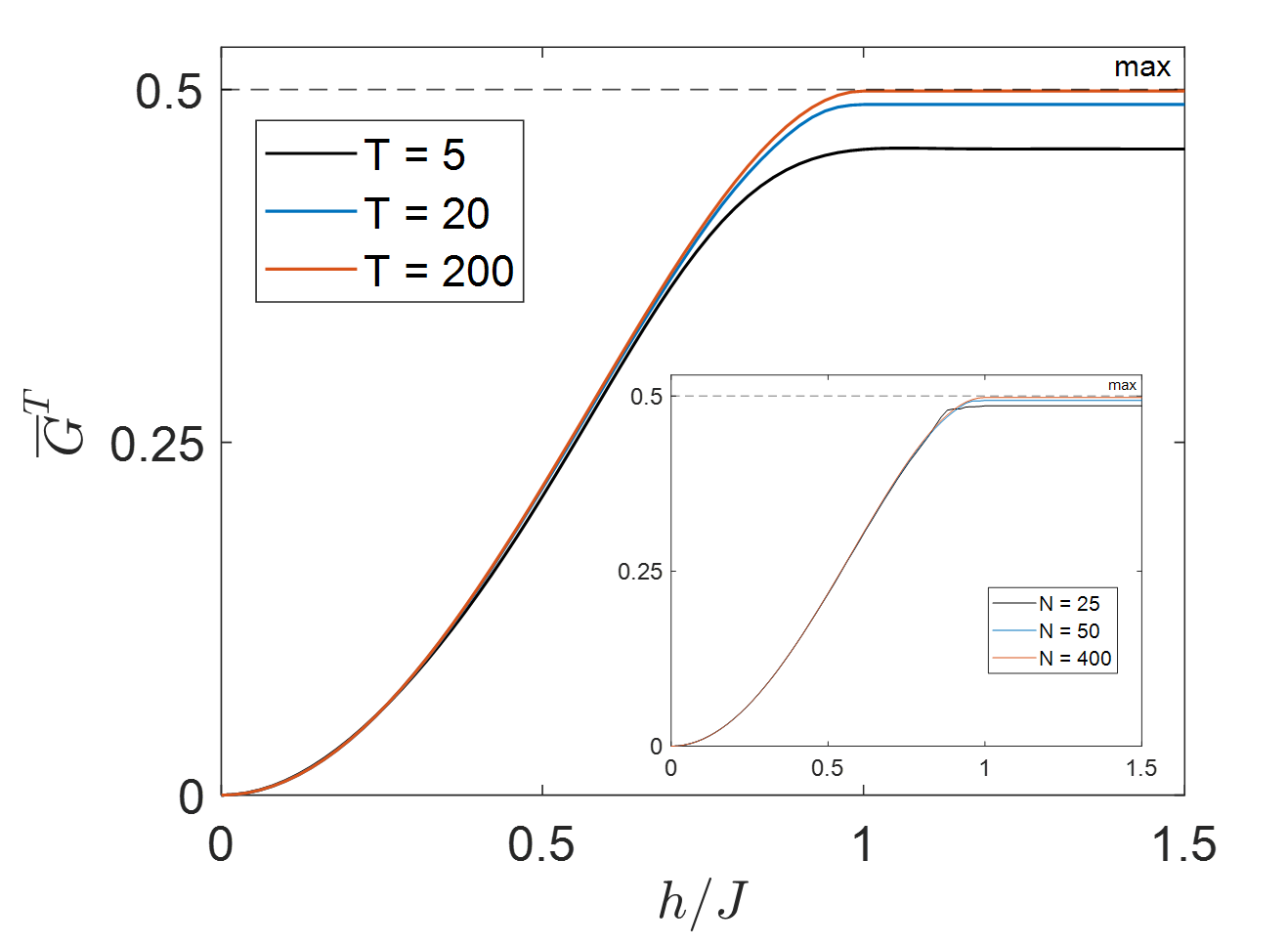}
  \caption{The long-time average of the A-OTOC for the edge-spin parity operator $\sigma^z_1$ in the TFIM (\ref{eq:TFIM}) calculated via numerical integration for a chain of size $N = 400$.  Here we use numerical integration so that the order of the long-time and thermodynamic limits are respected (i.e. for any finite size system, we must evaluate the long-time average of the A-OTOC via finite time integration, not the analytic expression given by Eq.~\ref{eq:bar{G}}), results are averaged over times $T = 5 - 200$.  The inset shows the A-OTOC averaged over $T = 100$ while varying system size from $N = 25 - 400$; the convergence of these plots shows that we reach the thermodynamic limit first for finite time averages $T\sim100$ (plots for $N = 200$ and $N = 400$ essentially overlap, ensuring the convergence).}
  \label{fig:order}
\end{figure}

It was shown in \cite{Kemp_2017} that the autocorrelation function for the edge spin operator $\langle s|\sigma^z_1(t)\sigma^z_1|s\rangle$ approaches $1-(\frac{h}{J})^2$ (where $|s\rangle$ is any eigenstate of the Hamiltonian) in the infinite time - thermodynamic limit, and thus captures the transition from order to disorder similarly to the ground state magnetization\footnote{Note that the limits fail to commute; the correct order to take the limits in this case is $\lim_{T\rightarrow\infty}\lim_{N\rightarrow\infty}$ since the observed order - disorder transition happens in the thermodynamic limit, and we are interested in the behavior of the long-time average autocorrelation function in this limit.}.  This quantity being the autocorrelation function of a parity operator suggests that the A-OTOC may also see this transition.  Indeed, Fig.~\ref{fig:order} shows the long-time average of the A-OTOC, and we see that it increases from 0 at $h=0$ to its ergodic value of $\approx\frac{1}{2}$ at the critical point, showing again the transition from order to disorder.

\subsection{Integrability Transition}
\begin{figure}[t]
  \centering
  \includegraphics[width=\columnwidth]{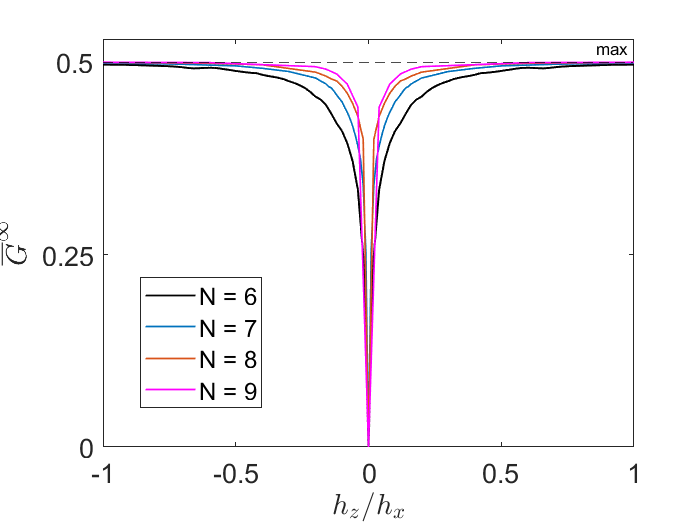}
  \caption{Infinte time average of the A-OTOC for the global parity operator $S=\bigotimes_i\sigma^x_i$ in the
  TFIM with a longitudinal field (\ref{eq:TFIM-L}) for different system sizes ($h_x=0.5$). Since $[S,H]=0$ at $h_z=0$,
  the long-time average $\overline{G}^{\,\infty}$ is exactly zero at this point,
  and it rapidly grows to $\approx\tfrac12$ for any $h_z\neq0$, producing a sharp drop in $\overline{G}^\infty$ that is symmetric in $h_z$. As the system size increases, we see that the drop in $\overline{G}^\infty$ begins to sharpen towards a singular point.}
  \label{fig:notch}
\end{figure}

Consider the Transverse-field Ising model (TFIM) in a longitudinal field.
\begin{equation}
    H = -\sum_{i=1}^{N-1}\sigma^z_i\sigma^z_{i+1} - h_x\sum_{i=1}^N\sigma^x_i-h_z\sum_{i=1}^N\sigma^z_i\label{eq:TFIM-L}
\end{equation}
At the integrable point $h_z=0$, the global $X$ parity operator $S=\bigotimes_{i=1}^N\sigma^x$ is a symmetry of the system, and it is broken by the presence of the longitudinal field.  Hence, at the integrable point, $\overline{G}^\infty$ is exactly 0, and any deviation from $h_z=0$ results in $\overline{G}^\infty$ rapidly approaching its ergodic value of $\approx\frac{1}{2}$ (since this model becomes chaotic as the longitudinal field is turned on).  Fig.~\ref{fig:notch} shows this quantity plotted for several different system sizes, and we see that the drop in $\overline{G}^\infty$ narrows as the system size increases since the strength of the integrability breaking field required for the onset of quantum chaos approaches 0 in the thermodynamic limit \cite{Bulchandani_2022}.

These examples show that the A-OTOC can capture a variety of phenomena exhibited by many-body systems.  The only modification required in each case is the choice parity operator that has structural significance to the model/phenomenon of interest.

\section{Conclusion\label{sec:conclu}}
In this paper, we continued the study of quantum scrambling for full algebras of observables started in \cite{Zanardi2022, Andreadakis2023}. We focused on algebras $\mathcal A$ whose commutant is generated by a Hermitian involution $S$, i.e., a ``parity’’ operator, and on closed-system dynamics.
The main tools employed for this study are the A-OTOC and the so-called Pl\"ucker fidelity.
The first tool is defined by an average measure of dynamically induced non-commutativity, while the second is a geometric measure of overlap between algebras suitably embedded in a higher-dimensional projective Hilbert space.

In this $\mathbb{Z}_2$ setting, both the A-OTOC and the Pl\"ucker fidelity admit simple closed
expressions that depend on the unitary defining the dynamics only through two scalars: the
trace of the $\mathbb{Z}_2$ generator $S$ and its infinite-temperature
dynamical autocorrelation function.
The two measures are tied by a simple algebraic identity, and both are governed by the
$\mathbb{Z}_2$-symmetry-breaking component of the evolution or, equivalently, by the transition probability between the two parity sectors.

We showed that these measures admit an elegant interpretation in terms of natural metric structures over the full manifold of $\mathbb{Z}_2$ algebras. In this way, the physical notion of quantum scrambling is tied, on the one hand, to the algebraic notion of non-commutativity and, on the other hand, to the geometric notion of distance between an algebra and its dynamical image.

We fully characterized the sets of non-scrambling and maximally scrambling unitaries, as well as the Haar-typical behavior in the $\mathbb{Z}_2$ setting. Typical unitaries or generators lead to maximal scrambling, and this property displays measure concentration in high dimensions.

For Hamiltonian channels, we derived the infinite-time average of the
autocorrelator and its temporal fluctuations under a non-resonance
assumption on the spectral gaps. The long-time scrambling is thus
controlled by the dephased component of the parity generator; after
further averaging over random eigenspaces, it is fixed by an entropic functional of
the degeneracy structure of the Hamiltonian, attaining its near-maximal
value for non-degenerate spectra.

Finally, we illustrated the formalism by applying it to one-dimensional spin chains featuring different phases, e.g., integrable, chaotic, and many-body localized phases.
By a judicious choice of the $\mathbb{Z}_2$ generator, the long-time averaged A-OTOC
may provide a sensitive diagnostic of the underlying phases and
of the symmetries that distinguish them.

Natural directions for future investigations aimed at assessing the scope and usefulness of the algebra-scrambling formalism include a more systematic and sophisticated numerical analysis of many-body systems in the $\mathbb{Z}_2$ case, as well as extensions to more general commutant types which are, however, still amenable to the rigorous mathematical analysis developed in this paper.
\begin{acknowledgments}
 PZ acknowledges partial support from the NSF award PHY2310227. The authors acknowledge useful discussion with F.  Iulianelli and E. Dallas.
\end{acknowledgments}
\bibliographystyle{apsrev4-2}
\bibliography{Z2-Refs.bib}

@article{Zanardi2022,
	author = {Zanardi, Paolo},
	doi = {10.22331/q-2022-03-31-666},
	journal = {Quantum},
	pages = {666},
	title = {Quantum scrambling of observable algebras},
	url = {https://doi.org/10.22331/q-2022-03-31-666},
	volume = {6},
	year = {2022}}

@article{Andreadakis2023,
	author = {Andreadakis, Fivos and Anand, Nikhil and Zanardi, Paolo},
	doi = {10.1103/PhysRevA.107.042217},
	journal = {Physical Review A},
	number = {4},
	pages = {042217},
	title = {Scrambling of algebras in open quantum systems},
	url = {https://doi.org/10.1103/PhysRevA.107.042217},
	volume = {107},
	year = {2023}}

@article{Zanardi2018,
	author = {Zanardi, Paolo and Campos Venuti, Lorenzo},
	doi = {10.1063/1.4997146},
	journal = {Journal of Mathematical Physics},
	number = {1},
	pages = {012203},
	title = {Quantum coherence generating power, maximally abelian subalgebras, and Grassmannian geometry},
	url = {https://doi.org/10.1063/1.4997146},
	volume = {59},
	year = {2018}}

@article{Zanardi2021,
	author = {Zanardi, Paolo and Anand, Nikhil},
	doi = {10.1103/PhysRevA.103.062214},
	journal = {Physical Review A},
	number = {6},
	pages = {062214},
	title = {Information scrambling and chaos in open quantum systems},
	url = {https://doi.org/10.1103/PhysRevA.103.062214},
	volume = {103},
	year = {2021}}

@article{Andreadakis2024,
	author = {Andreadakis, Fivos and Dallas, Ethan and Zanardi, Paolo},
	doi = {10.1103/PhysRevA.109.052424},
	journal = {Physical Review A},
	number = {5},
	pages = {052424},
	title = {Long-time quantum scrambling and generalized tensor product structures},
	url = {https://doi.org/10.1103/PhysRevA.109.052424},
	volume = {109},
	year = {2024}}

@article{Styliaris2021,
	author = {Styliaris, Georgios and Anand, Nikhil and Zanardi, Paolo},
	doi = {10.1103/PhysRevLett.126.030601},
	journal = {Physical Review Letters},
	number = {3},
	pages = {030601},
	title = {Information scrambling over bipartitions: Equilibration, entropy production, and typicality},
	url = {https://doi.org/10.1103/PhysRevLett.126.030601},
	volume = {126},
	year = {2021}}

@article{Zanardi2001_EntanglementQuantumEvolutions,
	author = {Paolo Zanardi},
	doi = {10.1103/PhysRevA.63.040304},
	journal = {Physical Review A},
	number = {4},
	pages = {040304(R)},
	title = {Entanglement of Quantum Evolutions},
	volume = {63},
	year = {2001}}

@article{ZanardiZalkaFaoro2000_EntanglingPower,
	author = {Paolo Zanardi and Christof Zalka and Lara Faoro},
	doi = {10.1103/PhysRevA.62.030301},
	journal = {Physical Review A},
	number = {3},
	pages = {030301(R)},
	title = {Entangling Power of Quantum Evolutions},
	volume = {62},
	year = {2000}}

@article{WangZanardi2002_QuantumEntanglementUnitaryOperators,
	author = {Xiaoguang Wang and Paolo Zanardi},
	doi = {10.1103/PhysRevA.66.044303},
	journal = {Physical Review A},
	number = {4},
	pages = {044303},
	title = {Quantum Entanglement of Unitary Operators on Bipartite Systems},
	volume = {66},
	year = {2002}}

@article{ProsenPizorn2007_OSEE_Ising,
	author = {Toma\v{z} Prosen and Iztok Pi\v{z}orn},
	doi = {10.1103/PhysRevA.76.032316},
	journal = {Physical Review A},
	number = {3},
	pages = {032316},
	title = {Operator Space Entanglement Entropy in a Transverse Ising Chain},
	volume = {76},
	year = {2007}}

@article{ZhouLuitz2017_OperatorEE_ChaoticSystems,
	author = {Tianci Zhou and David J. Luitz},
	doi = {10.1103/PhysRevB.95.094206},
	journal = {Physical Review B},
	number = {9},
	pages = {094206},
	title = {Operator entanglement entropy of the time evolution operator in chaotic systems},
	volume = {95},
	year = {2017}}

@article{PalLakshminarayan2018_EntanglingPower_ManyBody,
	author = {Rajarshi Pal and Arul Lakshminarayan},
	doi = {10.1103/PhysRevB.98.174304},
	journal = {Physical Review B},
	number = {17},
	pages = {174304},
	title = {Entangling power of time-evolution operators in integrable and nonintegrable many-body systems},
	volume = {98},
	year = {2018}}

@article{PizornProsen2009_OSEE_XY,
	author = {Iztok Pi\v{z}orn and Toma\v{z} Prosen},
	doi = {10.1103/PhysRevB.79.184416},
	journal = {Physical Review B},
	number = {18},
	pages = {184416},
	title = {Operator space entanglement entropy in {XY} spin chains},
	volume = {79},
	year = {2009}}

@article{MacCormackTanKudlerFlamRyu2021_NonthermalizingSystems,
	author = {Ian MacCormack and Mao Tian Tan and Jonah Kudler-Flam and Shinsei Ryu},
	doi = {10.1103/PhysRevB.104.214202},
	journal = {Physical Review B},
	number = {21},
	pages = {214202},
	title = {Operator and entanglement growth in nonthermalizing systems: Many-body localization and the random singlet phase},
	volume = {104},
	year = {2021}}

@article{LarkinOvchinnikov1969_QuasiclassicalMethod,
	author = {A. I. Larkin and Yu. N. Ovchinnikov},
	journal = {Soviet Physics JETP},
	number = {6},
	pages = {1200--1205},
	title = {Quasiclassical Method in the Theory of Superconductivity},
	volume = {28},
	year = {1969}}

@article{ShenkerStanford2014_BlackHolesButterfly,
	author = {Stephen H. Shenker and Douglas Stanford},
	doi = {10.1007/JHEP03(2014)067},
	journal = {Journal of High Energy Physics},
	pages = {067},
	title = {Black Holes and the Butterfly Effect},
	volume = {2014},
	year = {2014}}

@article{MaldacenaShenkerStanford2016_BoundOnChaos,
	author = {Juan Maldacena and Stephen H. Shenker and Douglas Stanford},
	doi = {10.1007/JHEP08(2016)106},
	journal = {Journal of High Energy Physics},
	pages = {106},
	title = {A Bound on Chaos},
	volume = {2016},
	year = {2016}}

@article{SwingleBentsenSchleierSmithHayden2016_MeasuringScrambling,
	author = {Brian Swingle and Gregory Bentsen and Monika Schleier-Smith and Patrick Hayden},
	doi = {10.1103/PhysRevA.94.040302},
	journal = {Physical Review A},
	number = {4},
	pages = {040302},
	title = {Measuring the Scrambling of Quantum Information},
	volume = {94},
	year = {2016}}

@article{ZanardiStyliarisCamposVenuti2017_CoherenceGeneratingPower,
	author = {Paolo Zanardi and Georgios Styliaris and Lorenzo Campos Venuti},
	doi = {10.1103/PhysRevA.95.052306},
	journal = {Physical Review A},
	number = {5},
	pages = {052306},
	title = {Coherence-generating power of quantum unitary maps and beyond},
	volume = {95},
	year = {2017}}

@article{AnandStyliarisKumariZanardi2021_CoherenceSignatureChaos,
	author = {Namit Anand and Georgios Styliaris and Meenu Kumari and Paolo Zanardi},
	doi = {10.1103/PhysRevResearch.3.023214},
	journal = {Physical Review Research},
	number = {2},
	pages = {023214},
	title = {Quantum coherence as a signature of chaos},
	volume = {3},
	year = {2021}}

@article{StyliarisAnandCamposVenutiZanardi2019_CoherenceLocalization,
	author = {Georgios Styliaris and Namit Anand and Lorenzo Campos Venuti and Paolo Zanardi},
	doi = {10.1103/PhysRevB.100.224204},
	journal = {Physical Review B},
	number = {22},
	pages = {224204},
	title = {Quantum coherence and the localization transition},
	volume = {100},
	year = {2019}}

@article{AnandZanardi2022_BROTOCs,
	author = {Namit Anand and Paolo Zanardi},
	doi = {10.22331/q-2022-06-27-746},
	journal = {Quantum},
	pages = {746},
	title = {{BROTOC}s and Quantum Information Scrambling at Finite Temperature},
	volume = {6},
	year = {2022}}

@article{StyliarisCamposVenutiZanardi2018_CoherenceGeneratingPowerDephasing,
	author = {Georgios Styliaris and Lorenzo Campos Venuti and Paolo Zanardi},
	doi = {10.1103/PhysRevA.97.032304},
	journal = {Physical Review A},
	number = {3},
	pages = {032304},
	title = {Coherence-generating power of quantum dephasing processes},
	volume = {97},
	year = {2018}}

@book{Ledoux2001Concentration,
  author    = {Ledoux, Michel},
  title     = {The Concentration of Measure Phenomenon},
  publisher = {American Mathematical Society},
  series    = {Mathematical Surveys and Monographs},
  volume    = {89},
  year      = {2001}
}

@article{PhysRevLett.111.127201,
  title = {Local Conservation Laws and the Structure of the Many-Body Localized States},
  author = {Serbyn, Maksym and Papi\ifmmode \acute{c}\else \'{c}\fi{}, Z. and Abanin, Dmitry A.},
  journal = {Phys. Rev. Lett.},
  volume = {111},
  issue = {12},
  pages = {127201},
  numpages = {5},
  year = {2013},
  month = {Sep},
  publisher = {American Physical Society},
  doi = {10.1103/PhysRevLett.111.127201},
  url = {https://link.aps.org/doi/10.1103/PhysRevLett.111.127201}
}

@misc{pawlik2026facetsbodylocalization,
      title={Facets of Many Body Localization}, 
      author={Konrad Pawlik and Maksym Prodius and Pedro R. Nicácio Falcão and Jakub Zakrzewski},
      year={2026},
      eprint={2601.09494},
      archivePrefix={arXiv},
      primaryClass={cond-mat.dis-nn},
      url={https://arxiv.org/abs/2601.09494}, 
}

@article{Bulchandani_2022,
   title={Onset of many-body quantum chaos due to breaking integrability},
   volume={105},
   ISSN={2469-9969},
   url={http://dx.doi.org/10.1103/PhysRevB.105.214308},
   DOI={10.1103/physrevb.105.214308},
   number={21},
   journal={Physical Review B},
   publisher={American Physical Society (APS)},
   author={Bulchandani, Vir B. and Huse, David A. and Gopalakrishnan, Sarang},
   year={2022},
   month=June }

@article{Kemp_2017,
	doi = {10.1088/1742-5468/aa73f0},
	url = {https://doi.org},
	year = 2017,
	month = {jun},
	publisher = {{IOP} Publishing},
	volume = {2017},
	number = {6},
	pages = {063105},
	author = {J Kemp and N Y Yao and N Y Lao and P Fendley},
	title = {Long coherence times for edge spins},
	journal = {Journal of Statistical Mechanics: Theory and Experiment}
}

@article{Pfeuty:1970qrn,
    author = "Pfeuty, Pierre",
    title = "{The one-dimensional Ising model with a transverse field}",
    doi = "10.1016/0003-4916(70)90270-8",
    journal = "Annals Phys.",
    volume = "57",
    number = "1",
    pages = "79--90",
    year = "1970"
}
\appendix
\section{Derivation of Eq.~(\ref{eq:outer})}
\noindent
{\em{Sufficient condition.}}
\medskip

\noindent
If $\mathcal{U}(S)=-S$, then $Y=-1$ and
\begin{align}
\tau=\frac{1}{d}\operatorname{Tr}(S)
=-\frac{1}{d}\operatorname{Tr}(\mathcal{U}(S))
=-\frac{1}{d}\operatorname{Tr}(S)
=-\tau . \nonumber
\end{align}
Therefore $\tau=0$ and $Y=2\tau^2-1$ is satisfied.

\bigskip

\noindent
{\em{Necessary condition.}}
\medskip

\noindent
If $S=\Pi^{+}-\Pi^{-}$, one has
\begin{align}
Y(U)
&=\frac{1}{d}\Bigl[
\operatorname{Tr}\bigl(\Pi^{+}\mathcal{U}(\Pi^{+})\bigr)
+\operatorname{Tr}\bigl(\Pi^{-}\mathcal{U}(\Pi^{-})\bigr) \nonumber\\
&\qquad\qquad
-\operatorname{Tr}\bigl(\Pi^{-}\mathcal{U}(\Pi^{+})\bigr)
-\operatorname{Tr}\bigl(\Pi^{+}\mathcal{U}(\Pi^{-})\bigr)
\Bigr], \nonumber
\end{align}
where
$\Pi^{\alpha}\equiv \operatorname{span}\Bigl\{
|\psi_k^{\alpha}\rangle
\Bigr\}_{k=1}^{d_\alpha},
\, (\alpha=\pm1). \nonumber
$

\noindent
Note
$\operatorname{Tr}\bigl(\Pi^{-}\mathcal{U}(\Pi^{+})\bigr)
=\sum_{k=1}^{d^-}
\langle \psi_k^{-}|\mathcal{U}(\Pi^{+})|\psi_k^{-}\rangle
\le d^- , \nonumber
$
and
\begin{align}
\operatorname{Tr}\bigl(\Pi^{-}\mathcal{U}(\Pi^{+})\bigr)
&=\operatorname{Tr}\bigl(\mathcal{U}^{\dagger}(\Pi^{-})\Pi^{+}\bigr) \nonumber\\
&=\sum_{k=1}^{d^+}
\langle \psi_k^{+}|\mathcal{U}^{\dagger}(\Pi^{-})|\psi_k^{+}\rangle
\le d^+ . \nonumber
\end{align}
The same works for $\operatorname{Tr}\bigl(\Pi^{+}\mathcal{U}(\Pi^{-})\bigr)$. Hence, for
$\alpha\neq\beta$,
\begin{align}
\operatorname{Tr}\bigl(\Pi^{\alpha}\mathcal{U}(\Pi^{\beta})\bigr)
\le \min\{d^{+},d^{-}\}. \nonumber
\end{align}

\noindent
On the other hand,
\begin{align}
\operatorname{Tr}\bigl(\Pi^{\alpha}\mathcal{U}(\Pi^{\alpha})\bigr)
&=\operatorname{Tr}\Bigl(\Pi^{\alpha}\bigl(\mathbb{I}-\mathcal{U}(\Pi^{\bar\alpha})\bigr)\Bigr)
\nonumber\\
&=d^{\alpha}-\operatorname{Tr}\bigl(\Pi^{\alpha}\mathcal{U}(\Pi^{\bar\alpha})\bigr)
\nonumber\\
&\ge d^{\alpha}-\min\{d^{+},d^{-}\},
\qquad (\bar\alpha=-\alpha). \nonumber
\end{align}
Thus
$Y(U)\ge {1-4d^{-1}\min\{d^{+},d^{-}\}}
.$

One can also write $d^{\pm}/d= (1\pm\tau)/2,$ from which 
\begin{equation}
Y(U)\ge 2|\tau|-1\ge 2\tau^2-1.\label{eq:Y-range}
\end{equation}
From here we see that $Y(U)=2\tau^2-1$ can be achieved for non-trivial $\mathbb{Z}_2$  just for $\tau=0,$
and when all the bounds above are saturated. The latter happens iff
$\mathcal{U}(\Pi^{\alpha})=\Pi^{\bar\alpha},
\, (\alpha=\pm1), \nonumber
$
which means
\begin{align}
\mathcal{U}(S)=\mathcal{U}(\Pi^{+})-\mathcal{U}(\Pi^{-})
=\Pi^{-}-\Pi^{+}
=-S . \nonumber
\end{align}
\section{Derivation of Eqs.~(\ref{eq:T-def}),(\ref{eq:GA-transition})}
We start by writing the autocorrelation function $Y$ as a function of the
projectors $f_\pm$. One has
\begin{align}
Y
&= \frac{1}{d}\operatorname{Tr}\bigl[S\,\mathcal{U}(S)\bigr]
 = \frac{1}{d}\operatorname{Tr}\bigl[(f_+-f_-)
 \mathcal{U}(f_+-f_-)\bigr] \\
&= \frac{1}{d}\operatorname{Tr}\bigl[(1-2f_-)
 \mathcal{U}(2f_+-1)\bigr] \\
&= \frac{1}{d}\operatorname{Tr}\bigl[2f_+-1
 -4 f_-\mathcal{U}(f_+) +2f_-\bigr] \\
&= 1-\frac{4}{d}\operatorname{Tr}\bigl[f_-\mathcal{U}(f_+)\bigr].
\end{align}
Now
$
\mathbb{E}_{\ket{\psi}\in\mathcal{H}_+}
\bigl[\ket{\psi}\!\bra{\psi}\bigr]
= \frac{f_+}{d^+},
$
whence 
$
Y=1-\frac{4d^+}{d}T(U),
$
which implies 
$
 \mathfrak{b}^2(U):=(1-Y)/{2}=2(d^+/{d})\,T(U).
$
But
$
 d^+=d(1+\tau)/2,$
therefore the first Eq.~(\ref{eq:T-def}) follows
\begin{equation}
 b^2(U)=(1+\tau)T(U).
\end{equation}
Plugging this result into Eq.~(\ref{eq:T-def}) one finds Eq.~(\ref{eq:GA-transition}). 

Note also (assume w.l.o.g.  $\tau\ge 0$)
\begin{equation}
T(U)=\frac{1}{d^+}\operatorname{Tr}\bigl[f_-\mathcal{U}(f_+)\bigr]
\leq \frac{d^-}{d^+}
=\frac{1-\tau}{1+\tau}
\leq 1-\tau .
\end{equation}
The first inequality is saturated iff
$
\mathcal{H}_-\subseteq \operatorname{Im}\mathcal{U}(f_+),
$
and the second if $\tau=0$. When both are valid,  one must then have
$
T(U)=1,$ and $
U\mathcal{H}_+=\mathcal{H}_- .$ This corresponds to outer unitaries swapping the two $\mathbb{Z}_2$-sectors and resulting in no-scrambling i.e., $G_{\mathcal A}=0.$

\section{Derivation of Eqs.~(\ref{eq:z2-HS}), (\ref{eq:metric-Z2})}
We first notice  that $\mathbb{P}_{\mathcal{U}(\mathcal{A}') }= \mathcal{U}\mathbb{P}_{\mathcal{A}'}  \mathcal{U}^\dagger.$ It follows that  the distance $D^2_{\mathrm{HS}}(\mathcal{A}',\mathcal{U}(\mathcal{A}') ),$ is given by
\begin{align}
\bigl\|\mathbb{P}_{\mathcal A'}&-\mathcal{U}\mathbb{P}_{\mathcal A'}\mathcal{U}^\dagger\bigr\|_{\mathrm{HS}}^2
=
\|\mathbb{P}_{\mathcal A'}\|_{\mathrm{HS}}^2
+\|\mathcal{U}\mathbb{P}_{\mathcal A'}\mathcal{U}^\dagger\|_{\mathrm{HS}}^2\nonumber \\
-&2\bigl\langle \mathbb{P}_{\mathcal A'},\mathcal{U}\mathbb{P}_{\mathcal A'}\mathcal{U}^\dagger\bigr\rangle
=
2\Bigl(d(\mathcal A')-\|\widetilde M_{\mathcal A}(\mathcal{U})\|_2^2\Bigr).
\end{align}

Here we used
\begin{align}
\|\mathbb{P}_{\mathcal A'}\|_{\mathrm{HS}}^2
&=
\operatorname{Tr}_{\mathrm{HS}}(\mathbb{P}_{\mathcal A'})
=
\sum_{\alpha=1}^{d(\mathcal A')}
\langle \tilde{f}_\alpha,\mathbb{P}_{\mathcal A'}(\tilde{f}_\alpha)\rangle
\nonumber\\
&=
\sum_{\alpha=1}^{d(\mathcal A')}
\|\tilde{f}_\alpha\|_2^2
=
d(\mathcal A')
=
\|\mathcal{U}\mathbb{P}_{\mathcal A'}\mathcal{U}^\dagger\|_{\mathrm{HS}}^2 .
\end{align}
and
\begin{align}
\operatorname{Tr}_{\mathrm{HS}}&\|\bigl[\mathbb{P}_{\mathcal A'}\,\mathcal{U}\mathbb{P}_{\mathcal A'}\mathcal{U}^\dagger\bigr]
=
\sum_{\alpha=1}^{d(\mathcal A')}
\Bigl\langle
\mathbb{P}_{\mathcal A'}(\tilde{f}_\alpha),
\mathcal{U}\mathbb{P}_{\mathcal A'}\mathcal{U}^\dagger(\tilde{f}_\alpha)
\Bigr\rangle \nonumber\\
&=
\sum_{\alpha=1}^{d(\mathcal A')}
\Bigl\langle
\mathbb{P}_{\mathcal A'}(\tilde{f}_\alpha),
\mathcal{U}\Bigl(
\sum_{\beta=1}^{d(\mathcal A')}
\tilde{f}_\beta\,\langle \tilde{f}_\beta,\mathcal{U}^\dagger(\tilde{f}_\alpha)\rangle
\Bigr)
\Bigr\rangle \nonumber\\
&=
\sum_{\alpha,\beta=1}^{d(\mathcal A')}
\langle \tilde{f}_\alpha,\mathcal{U}(\tilde{f}_\beta)\rangle\,
\langle \tilde{f}_\beta,\mathcal{U}^\dagger(\tilde{f}_\alpha)\rangle \nonumber\\
&=
\sum_{\alpha,\beta=1}^{d(\mathcal A')}
\bigl|\langle \tilde{f}_\alpha,\mathcal{U}(\tilde{f}_\beta)\rangle\bigr|^2
=
\|\widetilde M_{\mathcal A}(\mathcal{U})\|_2^2 .
\end{align}
Using Eq.~(\ref{eq:Mtilde}) one obtains
\begin{align}
\|\widetilde M_{\mathcal A}(\mathcal{U})\|_2^2
=
1+\left(\frac{Y-\tau^2}{1-\tau^2}\right)^2
=
1+F_{\mathcal A}^{\,2}(\mathcal{U}).
\end{align}

Finally, since $d(\mathcal A')=2$, one finds
\begin{align}
\frac{1}{2}\bigl\|\mathbb{P}_{\mathcal A'}-\mathcal{U}\mathbb{P}_{\mathcal A'}\mathcal{U}^\dagger\bigr\|^2
=
1-F_{\mathcal A}^{\,2}(\mathcal{U}).
\end{align}
Moreover, from $G_{\mathcal A}(U)=\langle \mathfrak{b}^2(U)\rangle_U (1-F^2_{\mathcal A}(U))$ and Eq.~(\ref{eq:dist-alg}) one immediately gets  (\ref{eq:z2-HS}),  and (\ref{eq:metric-Z2}).

\section{Derivation of Eq.~(\ref{eq:Y-W-ave})}
\begin{align}
Y_W
&=\frac{1}{d}\operatorname{Tr}\!\left[W S_0 W^\dagger U W S_0 W^\dagger U^\dagger\right] \nonumber\\
&=\frac{1}{d}\operatorname{Tr}\!\left[T\,W^{\otimes 2}S_0^{\otimes 2}W^{\dagger\otimes 2}(U\otimes U^\dagger)\right]. \nonumber
\end{align}

\begin{align}
\langle Y_W\rangle_W
&=\frac{1}{d}\operatorname{Tr}\!\left[T\,\Bigl\langle W^{\otimes 2}S_0^{\otimes 2}W^{\dagger\otimes 2}\Bigr\rangle_W (U\otimes U^\dagger)\right]. \nonumber
\end{align}
From the well-known identity for Haar 2-designs,
\begin{align}
\Bigl\langle W^{\otimes 2}S_0^{\otimes 2}W^{\dagger\otimes 2}\Bigr\rangle_W
&=\frac{1}{2d}\sum_{\alpha=\pm1}\frac{\mathbb{I}+\alpha T}{d+\alpha}\,
\langle \mathbb{I}+\alpha T,S_0^{\otimes 2}\rangle \nonumber\\
&=\frac{1}{2d}\sum_{\alpha=\pm1}\frac{\mathbb{I}+\alpha T}{d+\alpha}\,
(d^2\tau^2+\alpha d). \label{2-design}
\end{align}
Here $T$ is the swap operator in $\mathcal{H}^{\otimes 2}.$
Therefore,
\begin{align}
\langle Y_W\rangle_W
&=\frac{1}{2d}\sum_{\alpha=\pm1}
\left\langle \frac{T+\alpha\mathbb{I}}{d+\alpha},\,U\otimes U^\dagger\right\rangle
(d\tau^2+\alpha) \nonumber\\
&=\frac{1}{2d}\sum_{\alpha=\pm1}
\left(\frac{d+\alpha|\operatorname{Tr}U|^2}{d+\alpha}\right)
(d\tau^2+\alpha) \nonumber\\
&=1-(1-\tau^2)\left(\frac{d^2-|\operatorname{Tr}U|^2}{d^2-1}\right) \nonumber\\
&=1-\frac{1-\tau^2}{1-d^{-2}}
\left(1-\left|\frac{\operatorname{Tr}U}{d}\right|^2\right). \nonumber
\end{align}
Finally, by using the definition $$K(U):=\frac{|{\Tr(U)}|^2-1 }{d^2-1}$$ one gets  Eq.~(\ref{eq:Y-W-ave}).
\section{Derivation of Eqs.~(\ref{eq:conc-G}),(\ref{eq:conc-F})}
The result derives from Eq.~(13) of Ref.~\cite{Andreadakis2023}, by setting $d(\mathcal{A}')=2,$ and Eq.~(\ref{eq:D-ave}. For completeness we report here below 
an independent calculation specialized to the $\mathbb{Z}_2$-symmetric  case.

 From Eq.~(\ref{eq:G-F-Y}) we have
\begin{equation}
G_{\mathcal A}(U)=1-Y-\frac{(1-Y)^2}{2(1-\tau^2)}.
\end{equation}
From the previous equation we have $\langle Y(U)\rangle_U=\tau^2$. Therefore, to compute $\langle G(U)\rangle_U$, one needs only to evaluate the Haar average of
\begin{align}
Y^2(U)
&=\frac{1}{d^2}\bigl|\operatorname{Tr}[SUSU^{\dagger}]\bigr|^2 \nonumber\\
&=\frac{1}{d^2}\operatorname{Tr}\!\bigl[SUSU^{\dagger}\otimes SUSU^{\dagger}\bigr] \nonumber\\
&=\frac{1}{d^2}\operatorname{Tr}\!\bigl[S^{\otimes 2}U^{\otimes 2}S^{\otimes 2}U^{\dagger\otimes 2}\bigr].
\label{eq:Y2U}
\end{align}
Using again Eq.~(\ref{2-design}) and 
remembering,
$$\operatorname{Tr}S^{\otimes 2}=(\operatorname{Tr}S)^2=d^2\tau^2,\,\,\,
\operatorname{Tr}(S^{\otimes 2}T)=\operatorname{Tr}(S^2)=\operatorname{Tr}\mathbb{I}=d,$$
one finds
\begin{align}
\langle Y^2(U)\rangle_U
&=\frac{1}{2d^2}\sum_{\alpha=\pm 1}
\frac{\bigl| (\operatorname{Tr}S)^2+\alpha d \bigr|^2}{d(d+\alpha)} \nonumber\\
&=\frac{1}{2d^2}\left[
\frac{(d^2\tau^2+d)^2}{d(d+1)}+
\frac{(d^2\tau^2-d)^2}{d(d-1)}
\right] \nonumber\\
&=\tau^4+\frac{(1-\tau^2)^2}{d^2-1}.
\end{align}
From here we find the Haar variance of $Y(U)$
\begin{equation}
\sigma^2_U(Y):=  \langle Y^2(U)\rangle_U -\langle Y^(U)\rangle^2_U=\frac{(1-\tau^2)^2}{d^2-1}
\label{eq:var-Y}
\end{equation}
Finally,
\begin{align}
\langle G_{\mathcal A}(U)\rangle_U
&=1-\langle Y(U)\rangle_U
-\frac{1-2\langle Y(U)\rangle_U+\langle Y^2(U)\rangle_U}{2(1-\tau^2)} \nonumber\\
&=\frac{1-\tau^2}{2}\,\frac{d^2-2}{d^2-1}.
\end{align}
By concavity,
$\bigl\langle \sqrt{F_{\mathcal A}(U)}\bigr\rangle_U\leq \sqrt{\bigl\langle F_{\mathcal A}^2(U)\bigr\rangle}.
$
%
Since, from Eq.~(\ref{eq:G-F-Y}),
$F_{\mathcal A}(U)=(Y-\tau^2)(1-\tau^2)^{-1},
$
\begin{align}
\bigl\langle F_{\mathcal  A}^2(U)\bigr\rangle
&=\frac{\langle (Y-\langle Y\rangle_U)^2\rangle_U}{(1-\tau^2)^2}\nonumber \\ &=\frac{\sigma^2_U(Y)}{(1-\tau^2)^2}=\frac{1}{d^2-1}.
\end{align}
Therefore, one has the $\tau$-independent bound
\begin{equation}
\langle F_{\mathcal  A}(U)\rangle_U\leq (d^2-1)^{-1/2}.
\end{equation}
\section{The A-OTOC is Lipschitz}
This property can be shown in general directly from the definiton (\ref{eq:A-OTOC}).  Here we show the proof specialized to the $\mathbb Z_2$-symmetric case.
In this case we  start from
$$ 
G_{\mathcal A}(U)=\frac{1-\tau^2}{2}\bigl[1-F_{\mathcal A}(U)^2\bigr].
$$
Therefore,
\begin{align}
\bigl|G_{\mathcal A}(U)&-G_{\mathcal A}(W)\bigr|
\leq \frac{|1-\tau^2|}{2}
\bigl|F_{\mathcal A}(U)^2-F_{\mathcal A}(W)^2\bigr| 
\nonumber\\
&\leq \frac{|1-\tau^2|}{2}
\bigl(F_{\mathcal A}(U)+F_{\mathcal A}(W)\bigr)
\bigl|F_{\mathcal A}(U)-F_{\mathcal A}(W)\bigr|.\nonumber
\end{align}
Since $0\leq F_{\mathcal A}\leq 1$, this gives
\begin{align}
\bigl|G_{\mathcal A}(U)-G_{\mathcal A}(W)\bigr|
&\leq |1-\tau^2|
\bigl|F_{\mathcal A}(U)-F_{\mathcal A}(W)\bigr| 
\nonumber\\
&\leq |Y(U)-Y(W)| .
\end{align}
It remains to estimate the last term. One has
\begin{align}
|Y(U)-Y(W)|
&=\frac{1}{d}\left|\operatorname{Tr}\bigl(SUSU^\dagger\bigr)
-\operatorname{Tr}\bigl(SWSW^\dagger\bigr)\right| 
\nonumber\\
&\leq \left\|SUSU^\dagger-SWSW^\dagger\right\|_\infty 
\nonumber\\
&\leq \left\|US(U^\dagger-W^\dagger)\right\|_\infty
+\left\|(U-W)SW^\dagger\right\|_\infty 
\nonumber\\
&\leq 2\|U-W\|_\infty .
\end{align}
Finally,
\begin{equation}
\bigl|G_{\mathcal A}(U)-G_{\mathcal A}(W)\bigr|
\leq 2\|U-W\|_\infty
\leq 2\|U-W\|_2 .\label{eq:G-Lip}
\end{equation}
Here only standard operator inequalities have been used.

\section{Derivation of Eqs.~(\ref{eq:t-var}) and (\ref{eq:Y-ave})}
\noindent The basic long-time-average identities can also be written as follows. One has
$Y(t)=\operatorname{Tr}\bigl[S U_t S U_t^{\dagger}\bigr],
$
with $U_t=e^{-iHt}=\sum_m \Pi_m e^{-i\epsilon_m t}$
(spectral resolution). Then,
\begin{equation}
Y(t)=\sum_{m,n}\operatorname{Tr}\bigl[S\Pi_n S\Pi_m\bigr]e^{-i(\epsilon_n-\epsilon_m)t}.
\end{equation}
From
$\overline{e^{-i(\epsilon_n-\epsilon_m)t}}^{\,\infty}=\delta_{nm},
$
we get
\begin{align}
\overline{Y}^{\,\infty}
&=\sum_n \operatorname{Tr}\bigl(S\Pi_n S\Pi_n\bigr)
=\operatorname{Tr}\bigl[S\,D_H(S)\bigr],
\end{align}
where
$D_H(\cdot)=\sum_n \Pi_n\,\cdot\,\Pi_n
$
is the Hamiltonian dephasing superoperator. Since $D_H$ is a projector, one finally gets
\begin{equation}
\overline{Y}^{\,\infty}=\langle S,D_H(S)\rangle
=\langle D_H(S),D_H(S)\rangle
=\|D_H(S)\|_2^2.\nonumber
\end{equation}
Also, defining
$X_{nm}:=\operatorname{Tr}\bigl[S\Pi_n S\Pi_m\bigr]=X_{mn},
$
 one obtains
\begin{equation}
Y^2(t)=\sum_{nmpq}X_{nm}X_{pq}\,
 e^{-i(\epsilon_n-\epsilon_m+\epsilon_p-\epsilon_q)t}.\nonumber
\end{equation}
Assuming
$\epsilon_n-\epsilon_m+\epsilon_p-\epsilon_q=0$
iff either $\epsilon_n=\epsilon_m$ and $\epsilon_p=\epsilon_q$, or $\epsilon_p=\epsilon_m$ and $\epsilon_q=\epsilon_n$ (nonresonance condition), 
one can evaluate $\overline{\exp[-i(\epsilon_n-\epsilon_m+\epsilon_p-\epsilon_q)t]}^{\,\infty}$. One finds
\begin{align}
\overline{Y^2}^{\,\infty}
&=(\sum_m X_{mm})^2+\sum_{n\neq m}X_{nm}X_{mn} \nonumber\\
&=\bigl(\overline{Y}^{\,\infty}\bigr)^2+\sum_{n\neq m}X_{nm}^2.
\end{align}
The condition $n\neq m$ is to avoid overcounting the case $\epsilon_n=\epsilon_m=\epsilon_p=\epsilon_q$. The temporal variance is given by,
\begin{equation}
\overline{Y^2}^{\,\infty}-\bigl(\overline{Y}^{\,\infty}\bigr)^2
=\sum_{n\neq m}\bigl[\operatorname{Tr}(S\Pi_n S\Pi_m)\bigr]^2.
\end{equation}

\section{Numerical Methods}
For the MBL transition observed in the XXZ model and the integrability breaking observed in the TFIM, $\overline{G}^\infty$ is calculated via exact diagonalization of the Hamiltonian and the analytic expression for $\overline{G}^\infty$ given by (\ref{eq:bar{G}}).  

For the XXZ model, the Hamiltonian commutes with the total spin operator $S_{\textnormal{tot}}^{z}=\sum_{i=1}^N\sigma^z_i$, so the Hamiltonian conserves the total spin of any state (by total spin, we mean the number of 1's in any computational basis state).  Thus, by reordering the computational basis and grouping basis vectors by their total spin in the $z$ direction, the Hamiltonian becomes block diagonal.  Thus, we can diagonalize $H$ by diagonalizing it within each block; this allows computations involving larger spin-chians compared to simply diagonalizing the full $H$ since the blocks are size $\binom{N}{L}\times\binom{N}{L}$ where $L$ is the total spin of the basis states in a given block.  When the parity operator we are considering is also block diagonal in this reordered basis, we can compute the long-time average of the A-OTOC via matrix multiplication only within each block since
\begin{equation}
    \textnormal{Tr}(S\Pi_iS\Pi_j) = \sum_m\textnormal{Tr}\big(S^{(m)}\Pi_i^{(m)}S^{(m)}\Pi_j^{(m)}\big)
\end{equation}
where $S=\bigoplus_{m}S^{(m))}$ and $\Pi_i=\bigoplus_{m}\Pi_i^{(m)}$.  This works for the family of Pauli $z$ type parity operators $S_k=\prod_{i=1}^k\sigma_i^z$ considered in the main text.  However, it fails for the random parity operators; in this case we evaluate $\textnormal{Tr}(S\Pi_iS\Pi_j)$ by matrix multiplication on the full Hilbert space.

At 0 disorder the system exhibits many degeneracies in its spectral gaps, so we evaluate $\overline{G}^\infty$ numerically via the "gap-resolved' expression that is valid when the non-resonance condition fails.  Since calculating the gaps is numerically costly, for all non-zero levels of disorder we calculate $\overline{G}^\infty$ assuming that the gaps in the spectrum are non-degenerate.  This is because in the presence of even small disorder, the XXZ model exhibits spectral statistic consistent with Wigner-Dyson \cite{pawlik2026facetsbodylocalization}, so it will satisfy this condition to good approximation.

For the plots with $S_k = \prod_{i=1}^k\sigma^z_i$ parity operators, we average $\overline{G}^\infty$ over 20 relizations of the disordered field with $N = 10$, and for the plots with $\sigma^z_1$ for different system sizes, we average over 250,200,200,20 realizations for $N = 7,8,9,10$, respectively. In each of these cases, for a given parity operator, the disorder average at all values of $W$ is calculated using the same random field, but scaled by the current value of W.  All error bars are to 2 standard errors.  

For the random parity operators, we want to average over not only the random fields in the model, but also over the random unitaries we use to generate the parity operators $S = Vs_AV^\dagger\otimes I_{B}$.  Essentially, we are interested in the quantity
\begin{equation}
    \mathbb{E}_{V,D}\;\overline{G}^\infty(V,D) 
\end{equation}
where $V$ is unitary and $D$ the vector whose elements are the random fields in the model.  The average is over the uniform distribution for $D$ and the Haar distribution for $V$.  We approximate this average by calculating $\overline{G}^\infty$ for 1000 pairs random unitaries $V$ and fields $D$.  For each line plot, the same 1000 unitaries and random fields (scaled by the current value of $W$) are used. The error bars are to two standard errors.

For the TFIM in a longitudinal field, the Hamiltonain fails to commute with a total spin operator, so we resort to exact diagonalization of the full Hamiltonain.  For $h_z\neq0$, $\overline{G}^\infty$ is computed via the assumption of non-degenerate spectral gaps; at $h_z=0$ we simply set $\overline{G}^\infty=0$. 

For the clean TFIM, we can perform computations with larger systems by first transforming the system by introducing Jordan-Wigner Majorana operators 
\begin{equation}
    a_j := \left(\prod_{k<j}\sigma^x_k\right)\sigma^z_j,\; b_j := \left(\prod_{k<j}\sigma^x_k\right)\sigma^y_j,\;j=1,\dots,N.
\end{equation}
The operators satisfy the anticommutation relations
\begin{equation}
    \{a_j,a_k\} = 2\delta_{jk},\; \{b_j,b_k\} = 2\delta_{jk},\; \{a_j,b_k\} = 0.
\end{equation}
This allows us to express the TFIM Hamiltonian as 
\begin{equation}
    H = -ih\sum_{j=1}^Na_jb_j - iJ\sum_{i=1}^{N-1}b_ja_{j+1}
\end{equation}
We can express this more compactly by defining $\gamma:=(a_1,b_1,\dots,a_Nb_N)^T$; the Hamiltonian them becomes
\begin{equation}
    H = \frac{i}{4}\gamma^TM\gamma,
\end{equation}
where $M$ is a $2N\times 2N$ real, antisymmetric matrix whose only non-zero entries are given by 
\begin{gather}
\begin{split}
    M_{2j-1,2j} = -2h,\;M_{2j,2j-1} = 2h,\;J=1,\dots,N,\\
    M_{2j,2j+1} = -2J,\;M_{2j+1,2j} = 2J,\;J=1,\dots,N-1.
\end{split}
\end{gather}
It them follows from the Heinsenberg equation of motion that $\dot{\gamma} = M\gamma\;\Rightarrow\;\gamma = e^{tM}\gamma$.  From this we then get that the edge-spin autocorrelation function is $Y(t)=\frac{1}{d}\textnormal{Tr}(\sigma^z_1\sigma^z_1(t))=(e^{tM})_{1,1}$.  We can then numerically integrate this function for large system size since we only need to exponentiate a $2N\times2N$ matrix.


\end{document}